\documentclass[a4paper,11pt]{article}
\pdfoutput=1
\usepackage{booktabs}
\usepackage{graphicx}
\usepackage{epsf,amsmath,bbold,amsfonts,stmaryrd}
\usepackage[utf8]{inputenc}
\usepackage{mathrsfs}
\usepackage{appendix}
\usepackage{amssymb}
\usepackage{float}
\usepackage{color}
\usepackage{cite}
\usepackage{hyperref}
\hypersetup{pageanchor=false}
\usepackage{indentfirst}
\usepackage{url}
\usepackage{float}
\usepackage{caption}
\usepackage[numbers,square,comma,sort&compress,merge]{natbib}
\usepackage{lineno}

\begin{document}

\title{\textbf{Screen chaotic motion by Shannon entropy in curved spacetimes}}
\author{Wenfu Cao$^{1\ast}$,
Yang Huang$^{1\dagger}$,
and
Hongsheng Zhang$^{1\ddagger}$}
\date{}

\maketitle

\vspace{-10mm}

\begin{center}
{\it
$^1$School of Physics and Technology, University of Jinan,
336 West Road of Nan Xinzhuang, Jinan, Shandong 250022, China\\\vspace{1mm}}
\end{center}

\vspace{8mm}

\begin{abstract}
We find a novel characteristic for chaotic motion by introducing Shannon entropy for periodic orbits, quasiperiodic orbits, and chaotic orbits. We compare our approach with the previous methods
including Poincaré Section, Lyapunov exponent, Fast Lyapunov Indicator, Recurrence plots (Rps), and Fast Fourier Transform (FFT) for orbits around black holes immersed in magnetic fields, and show
that they agree with each other quite well. The approach of Shannon entropy is intuitively clear, and theoretically reasonable since it becomes larger and larger from a periodic orbit to chaotic
orbit. We demonstrate that Shannon entropy can be a powerful probe to distinguish between chaotic and regular orbits in different spacetimes, and reversely may lead to a new route to define the
entropy for a single orbit in phase space, and to find more fundamental relations between thermodynamics and dynamics. Furthermore, we find that the fluctuations of entropy of chaotic orbits are
stronger than those of order orbits.
\end{abstract}

\vfill{\footnotesize Email: 202411100001@stu.ujn.edu.cn,\,\,sps\_Huangy@ujn.edu.cn,\,sps\_zhanghs$@$ujn.edu.cn,\\$~~~~~~\ddagger$ Corresponding author.}

\maketitle

\newpage

\section{Introduction}
Entropy is a central concept in physics and related fields. From the proposition of the concept of entropy by Clausius at 1850, entropy has been introduced in various ways and in various forms.
Entropy naturally describes the complexity of a system, especially a system with particles of macro quantity. Von Neumann shows that one can define an entropy for a single particle/state at quantum
level. In the distinguishable particle/mixed state limit, the Von Neumann entropy degenerates to Shannon entropy, which is the key concept in information theory, introduced by Shannon in the
1940s \cite{Shannon:1948,Parrondo:2015}.
It quantifies the uncertainty or randomness of information. The essence of information lies in its ability to reduce uncertainty, and Shannon entropy precisely measures this uncertainty.
Specifically, Shannon entropy quantifies the uncertainty of any informative dataset, including physical systems, by considering the probability distribution of all possible outcomes of an event.
Moreover, Shannon entropy can be viewed as a candidate for non-equilibrium entropy, and combining it with orbital dynamics may provide insights into dynamical entropy. Because information is
ubiquitous in almost all branches of sciences including physics, chemistry, biology etc., and even humanity,  Shannon entropy transcends the scope of communication theory or statistical physics, and
has been applied several other areas. We shall extend the concept of Shannon entropy to orbits, in especial the chaotic orbits surrounding black holes.

In recent years, the application of Shannon entropy has expanded from classical physical systems to the field of astronomy.
Early studies proposed Shannon entropy as a tool for modulating periodic orbits \cite{Li:1995}, with the core idea being that the Shannon entropy of periodic signals is proportional to their
complexity. Specifically, the entropy of a completely regular periodic signal is zero, while the entropy of a chaotic signal is higher. Additionally, studies have shown that chaos diffusion
estimators based on Shannon entropy can more accurately measure the diffusion range and diffusion rate of a set of orbits \cite{Giordano:2018}. Shannon entropy not only reflects the degree of
diffusion of a given initial set of orbits, but also, through related time derivative indicators, can effectively estimate the diffusion rate. This approach has been widely applied in the study of
dynamical stability in  multi-body systems \cite{AlvesSilva:2021,Kovari:2022}. Analysis of specific near-resonant systems, such as HD181433, has shown that Shannon entropy can complement other
dynamical indicators. In the Kepler-60 exoplanet system, Shannon entropy revealed the phase space structure and quantitatively described the chaotic diffusion process and planetary stability. These
studies further demonstrate the potential of Shannon entropy as a numerical tool for providing information on chaotic diffusion and dynamical stability in high-dimensional systems.

The chaotic phenomena in Newtonian mechanics have been extensively investigated‌. Within the framework of Newtonian mechanics, spacetime itself has no thermodynamic properties, thereby excluding concepts such as energy, temperature, and entropy‌. However, general relativity fundamentally alters this perspective by ascribing thermodynamic characteristics to spacetime, exemplified by black hole thermodynamics \cite{BHther1, BHther2}, cosmological thermodynamics \cite{COSther}, and even thermodynamic derivations of gravitational field equations \cite{ThertoEq1, ThertoEq2}‌. Consequently, thermodynamic laws—particularly the entropy increase law—must account for both matter-field entropy and gravitational-field entropy due to their mutual transformation‌.

In classical physics, defining entropy for few-body systems presents inherent challenges‌. This complexity underscores the physical significance of employing information-theoretic approaches to study orbital entropy in curved spacetime, especially the entropy of individual trajectories‌. However, it is important to note that in the study of dynamics in curved spacetime, the high complexity of the geodesic equations makes it difficult to distinguish between quasi-periodic and chaotic orbits based solely on the magnitude or variation of entropy. Therefore, it becomes crucial to further explore the properties of Shannon entropy in dynamical systems in curved spacetimes. From a
thermodynamic perspective, Shannon entropy encodes the information of geodesics in curved spacetime, imbuing spacetime with probabilistic features. At the same time, this characteristic may provide
new insights into the relationship between spacetime geometry and thermodynamics.

Black hole, as a strong gravitational source, whose existence is convinced by the direct imaging of black
holes\cite{Akiyama:2019cqa,Akiyama:2019brx,Akiyama:2019sww,Akiyama:2019bqs,Akiyama:2019fyp,Akiyama:2019eap} and the detection of gravitational waves\cite{LIGOScientific:2016aoc}, has several fancy
and unprecedent properties. The complex and bizarre orbits of null and time-like particles captured by the terrestrial and spatial apparatuses, for example the EHT, can be used to study the
properties black hole spacetime and furthermore,  the underlying gravitational theory.

Magnetic fields play a crucial role in the  astrophysical processes surrounding black holes, which are traditionally believed to be generated by accretion disks.
Recent observations imply that strong magnetic fields exist around black holes even in the absence of accretion disks\cite{Eatough:2013nva}. This suggests that black holes may be generally immersed
in  large-scale external electromagnetic fields. In the case of magnetars, these large-scale magnetic fields exhibit dipole characteristics but can be approximated as asymptotically uniform fields
in regions distant from the source\cite{Kovar:2014tla,{Stuchlik:2015nlt}}.

In this paper we concentrate on dynamics of particles around black holes immersed in asymptotically uniform magnetic fields, known as the Wald solution\cite{Wald:1974np}, and especially endow
Shannon entropy for the orbits of these surrounding particles.
Note the Wald potential is valid only when spacetime satisfies Ricci vacuum conditions, for spacetimes with modified gravity, appropriate adjustments should be considered\cite{Azreg-Ainou:2016tkt}.

When a black hole is immersed in an external magnetic field, the spacetime possesses only three conserved quantities: the energy of charged particles, angular momentum, and mass. The absence of a
fourth integral of motion renders the system non-integrable, leading to the emergence of chaotic dynamics \cite{Karas:1992,Takahashi:2008zh}.
The previous tools for identifying chaos include the Poincar\'{e} map, Lyapunov exponents and power spectra\cite{Gerald:1988,Li:2018wtz,Zhang:2023lrt}.
Recently, several  chaos indicators have been applied to the field of particle dynamics within relativistic gravity theory.
Recurrence plots(RPs), similar to the Poincar\'{e} map, can visualize phase space\cite{Kopacek:2010yr,Cao:2024ihv}.
The Lyapunov indicator in curved spacetime is a generalization of the Lyapunov exponent from flat spacetime\cite{Wu:2003pe,Wu:2006rx}.
With the aid of these chaos identification tools, we have established a preliminary result into the chaos generation mechanisms for charged and neutral particles near black holes.
When conserved energy $E$ is increased to high levels or in strong magnetic fields, particles generally exhibit chaotic phenomena\cite{Li:2018wtz,Hu:2021gwd,Yi:2020shw}.
Nevertheless, the role of black hole spin $a$ in inducing chaos is very intricate\cite{Kopacek:2010yr,Sun:2021ndd}.

The paper is organized as follows. The section \ref{sec:two} reviews the criteria for assessing the integrability of systems.
In section \ref{sec:three} we calculate the generalized Wald potentials for black holes in modified gravity.
The section \ref{sec:four} briefly introduces the application of symplectic algorithms in general relativity.
The section \ref{sec:five} describes our novel approach for calculating Shannon entropy of the orbits.
In section\ref{sec:six}, we demonstrate the power of Shannon entropy to identify the orbital states of particles moving in different spacetimes.
Finally, the section \ref{sec:seven} summarizes the analysis results.

\section{Integrability of spherically symmetric spacetimes}\label{sec:two}
Spherically symmetric spacetimes are typically considered integrable,
assuming that the black hole's internal electromagnetic field is ignored or the
black hole is not affected by an external magnetic field.

In this context, we assume that a simple spherically symmetric spacetime takes the following form
\begin{eqnarray}
ds^{2} &=&-A(r)dt^{2}+B(r)dr^{2}+C(r)d\Omega^{2}\nonumber \\
&=&-(1-\frac{f_{1}}{r}+\frac{f_{2}}{r^2})dt^{2}
+(1-\frac{f_{1}}{r}+\frac{f_{2}}{r^2})^{-1} dr^{2}\nonumber\\
&&+r^{2}( d\theta^{2} +\sin^{2} \theta d \varphi^2),
\end{eqnarray}
where $f_{1}$ and $f_{2}$ are independent of the coordinates $\{t,r,\theta,\varphi\}$.

The black hole solution has one or two  horizons
$r_{\pm}=\frac{f_{1}\pm\sqrt{f_{1}^{2}-4f_{2}}}{2}$, when $f_{1}=2M$ and $f_{2}=0$,
this reduces to the Schwarzschild case.
The spacetime has two Killing vectors, $\xi^{a}_{t}=(1,0,0,0)$ and
$\xi^{a}_{\varphi}=(0,0,0,1)$, which correspond to the specific energy
$E$ and orbital angular momentum  $L$, respectively. The specific energy and
angular momentum for test particles are given by
\begin{eqnarray}
E&=&-u^{a}\xi_{t a}=-u^{a}g_{ab}\xi^{b}_{t} =-g_{tt}\frac{dt}{d\tau}=-p_t, \label{eq:11}\\
L&=&u^{a}\xi_{\varphi a}=u^{a}g_{ab}\xi^{b}_{\varphi} = g_{\varphi\varphi}\frac{d\varphi}{d\tau}=p_{\varphi}, \label{eq:12}
\end{eqnarray}
where $\tau$ is the proper time of the particle. For convenience, throughout this paper we adopt geometric units in which the gravitational constant $G$, the black hole mass $M$, the speed of light
$c$, and the mass of the test particle $m$ are all set to unity, i.e., $G=c=M=m=1$.
The dynamics of a test particle with mass $m$ orbiting a black hole are governed by a Hamiltonian system
\begin{eqnarray}
H&=&\frac{1}{2} g^{ab}p_{a}p_{b} \nonumber \\
&=&-\frac{E^2}{2 \left(1-\frac{f_{1}}{r}+\frac{f_{2}}{r^2}\right)}+\frac{L^2 \csc^2\theta}{2 r^2} \nonumber \\
&& +\frac{p^2_r}{2} \left(1-\frac{f_{1}}{r}+\frac{f_{2}}{r^2}\right)+\frac{p_{\theta }^2}{2 r^2}.
\end{eqnarray}
This Hamiltonian is a conserved quantity for time-like particles,
\begin{equation}
H=-\frac{1}{2}.
\end{equation}
At this stage, we have determined three conserved quantities in the spacetime: the particle's mass,
energy, and angular momentum. If a fourth conserved quantity exists, the system would be completely
integrable. By separating the variables in the Hamilton-Jacobi equation for the Hamiltonian system(4),
the fourth conserved quantity $C_{k}$ reads
\begin{eqnarray}
C_{k}&=& \frac{r^2E^2}{ \left(1-\frac{f_{1}}{r}+\frac{f_{2}}{r^2}\right)}-r^2
-r^2p^2_r \left(1-\frac{f_{1}}{r}+\frac{f_{2}}{r^2}\right) \nonumber \\
&=&p_{\theta }^2+L^2 \csc^2\theta,
\end{eqnarray}
where $C_k$ is the Carter-like constant.
Thus, the Hamiltonian system (4) is integrable and has formally analytical
solutions.

\section{The external magnetic field of spherically symmetric black holes}\label{sec:three}

Wald\cite{Wald:1974np} studied the external magnetic field of vacuum, static, axial symmetric
black holes by means of Killing vector fields. It satisfies the following equation
\begin{equation}
\pounds_{\xi}g_{ab}=\nabla_{a}\xi_{b}+\nabla_{b}\xi_{a},
\end{equation}
where $\pounds_{\xi}g_{ab}$ denotes the Lie derivative of the tensor field $g_{ab}$
along the Killing vector field $\xi^{a}$.
Starting from the definition of the curvature tensor $R_{abc}^{\;\;\;\;\;d}$ and equation (7),
a very useful relation can be obtained
\begin{equation}
\nabla^{a}\nabla_{a}\xi^{b}=-R^{b}_{~d}\xi^{d},
\end{equation}
On the other hand, in curved space-time, a similar expression also exists for the source-free Maxwell equations with respect to a vector potential $A_{a}$ in the Lorentz gauge, which are given by
\begin{equation}
\nabla^{a}\nabla_{a}A^{b}=R^{b}_{~d}A^{d}.
\end{equation}
It is straightforward to see that Eqs. (8) and (9) have the same form. Subtracting Eq. (8) from Eq. (9) yields
\begin{equation}
\nabla^{a}\nabla_{a}(A^{b}-\xi^{b})=R^{b}_{~d}(A^{d}+\xi^{d}).
\end{equation}
In the case of a vacuum space-time, where
$R_{ab}=0$, the Killing vector field can serve as a solution to the source-free Maxwell equations in the Lorentz gauge.
This means that, the Killing vector field $\xi^{a}$ in vacuum must differ from the vector potential $A^{a}$ by a constant, and can be used to describe the solution for an external magnetic field
along the axis of symmetry in a static, axial symmetric spacetime. Therefore, the vector potential can be written as
\begin{equation}
A^{a}=C_{1}\xi_{(t)}^{a}+C_{2}\xi_{(\varphi)}^{a},
\end{equation}
where coefficients $C_{1}$ and $C_{2}$ depend on the magnetic field strength $B$.
For an axisymmetric, uncharged black hole, the coefficients obtained from the Wald's solution are
\begin{equation}
C_{1}=aB, ~~~~ C_{2}=\frac{B}{2},
\end{equation}
where $a$ is the angular momentum per unit mass of the black hole.
Azreg-A\"{i}nou \cite{Azreg-Ainou:2016tkt}  extended this result to more general non-vacuum scenarios, including
charged black holes and modified gravity black holes. In both cases, the modified potential
solutions must satisfy the source-free Maxwell equations
\begin{equation}
\nabla_{a}F^{ab}=0, \nabla_{[a}F_{bc]}=0.
\end{equation}
This paper focuses solely on spherically symmetric, uncharged black holes,
with the space-time metric given by Eq. (1). Following Azreg-A\"{i}nou's idea,
we rewrite Eq. (11) as
\begin{equation}
A^{a}=C_{1}(r,B)\xi_{(t)}^{a}+C_{2}(r,B)\xi_{(\varphi)}^{a},
\end{equation}
where $C_{2}=c_{2}(r,B)+\frac{B}{2}$. Thus, the two non-zero components
$(J_{t},J_{\varphi})$ of the source-free Maxwell equations are
\begin{eqnarray}
J_{t}&=&r^2 \left(-f_1 r+f_2+r^2\right) c_1''(r) \nonumber \\
&&+2 r \left(r^2-f_2\right) c_1'(r)+2 f_2 c_1(r)=0, \\
J_{\varphi}&=&r^2 \left(-f_1 r+f_2+r^2\right) c_2''(r)\nonumber \\
&&r^2 \left(4 r-3 f_1\right) c_2'(r)-2 f_2 \left(c_2(r)-r c_2'(r)\right)=0,
\end{eqnarray}
where the prime denotes derivative with respect to $r$. By specifying the boundary conditions for the above equations
such that the $C_{1}$ term resembles the Coulomb potential and the term $C_{2}$  reduces to the Wald potential
at infinity, we arrive at the exact solution for uncharged black holes
\begin{equation}
A_{a}=\frac{1}{2} B \left(r^2-f_2\right) \sin ^2(\theta )\xi^{(\varphi)}_{a}.
\end{equation}
We next present in Table 1 the generalized Wald potential for black holes and provide constraints on model parameters\cite{Vagnozzi:2022moj,Wang:2022yvi}.
References\cite{Cao:2024ihv,Abdujabbarov:2008mz,Abdujabbarov:2011uc} also provide solutions for the electromagnetic potential of other types of black holes that strictly satisfy the source-free
vacuum Maxwell equations. With this generalized Wald potential, we will be able to study the motion of charged particles very accurately near the black hole event horizon.

\section{Explicit symplectic algorithms for hamiltonian systems} \label{sec:four}
The charged particle motion in curved spacetime is described by\cite{Kopacek:2010yr}
\begin{eqnarray}
\cal{H} &=& \frac{1}{2}g^{ab}(P_{a}-qA_{a})(P_{b} -qA_{b})
\nonumber \\
&=& -\frac{1}{2}\left(1-\frac{f_{1}}{r}+\frac{f_{2}}{r^2}\right)^{-1}
E^{2} \nonumber \\
&& +\frac{1}{2}\left(1-\frac{f_{1}}{r} +\frac{f_{2}}{r^2}\right)p^{2}_{r}
+\frac{1}{2}\frac{p^{2}_{\theta}}{r^2} \nonumber \\
&& +\frac{1}{2r^2\sin^2\theta}\left[L-\frac{b}{2}\left(r^2-f_{2}\right)\sin^{2}\theta\right]^{2},
\end{eqnarray}
where $b=qB$, $P_{a}$ is the canonical four-momentum, which is related to the kinematical four-momentum
$p_{a}$ by the relation
 \begin{eqnarray}
P^{a}=p^{a}+qA^{a}.
\end{eqnarray}
When a black hole is immersed in an external magnetic field, the Hamiltonian equations cannot be separated to yield an expression like Eq. (6). This indicates the absence of a fourth constant of
motion, making the system non-integrable. Consequently, reliable numerical methods are necessary to study the dynamics of charged particles.

Symplectic schemes are considered the most suitable solvers for long-term integration of Hamiltonian systems, as they preserve the symplectic structure inherent in Hamiltonian dynamics. Recent
literature has focused on constructing explicit symplectic schemes using the multi-part splitting method\cite{Wang:2021gja,Wu:2021rrd,Zhou:2022uht}.
The Hamiltonian (18) can be separated in the form
\begin{equation}
\cal{H}=\cal{H}_1+\cal{H}_2+\cal{H}_3+\cal{H}_4+\cal{H}_5,
\end{equation}
where all sub-Hamiltonians are written as follows:
\begin{eqnarray}
\cal{H}_1 &=& \frac{1}{2r^2\sin^2\theta}\left[L-\frac{b}{2}(r^2-f_{2})\sin^{2}\theta\right]^{2} \nonumber
\\ && -\frac{{E}^{2}}{2}\left(1-\frac{f_{1}}{r}+\frac{f_{2}}{r^2}\right)^{-1}, \\
\cal{H}_{2} &=& \frac{1}{2}p^{2}_{r},\\
\cal{H}_{3} &=& -\frac{f_{1}}{2r}p^{2}_{r},\\
\cal{H}_{4} &=& \frac{1}{2r^2}p^{2}_{\theta}, \\
\cal{H}_{5} &=& \frac{f_{2}}{2r^2}p^{2}_{r}.
\end{eqnarray}
It is straightforward to verify that each of the five parts has an analytical solution
expressed as an explicit function of the proper time $\tau$. Solvers for the sub-Hamiltonians  $\cal{H}_1$, $\cal{H}_2$, $\cal{H}_3$, $\cal{H}_4$, and $\cal{H}_5$ are termed $\varkappa_1$,
$\varkappa_2$, $\varkappa_3$, $\varkappa_4$ and $\varkappa_5$, respectively.

Setting $h$  as the time step, we define a second-order explicit symplectic integrator as follows
\begin{eqnarray}
S2 (h)= \chi^*(\frac{h}{2})\times \chi(\frac{h}{2}),
\end{eqnarray}
where the two first-order solvers are given by
\begin{eqnarray}
    \chi(h) &=&\varkappa_5(h)\times \varkappa_4(h)\times \varkappa_3(h)\times \varkappa_2(h)\times \varkappa_1(h), \\
    \chi^*(h) &=& \varkappa_1(h)\times \varkappa_2(h)\times \varkappa_3(h)\times \varkappa_4(h)\times \varkappa_5(h).
\end{eqnarray}
By composing three second-order methods, we obtain a fourth-order explicit symplectic algorithm given by
\begin{eqnarray}
    S4=S2(\gamma h)\times S2(\delta  h)\times S2(\gamma h),
\end{eqnarray}
where $\gamma =1/(1-\sqrt[3]{2}) $ and $\delta =1-2\gamma$.
The construction follows that of Yoshida\cite{Yoshida:1990zz}. By utilizing components of higher-order operators  $\chi$ and $\chi^*$, an optimized fourth-order  partitioned
Runge-Kutta (PRK) symplectic algorithm was presented in\cite{Zhou:2022uht} as follows
\begin{eqnarray}
PRK_64 &=& \chi^* (\alpha _{12} h)\times \chi(\alpha _{11}
h)\times \cdots \nonumber \\ && \times \chi^* (\alpha _2 h) \times
\chi(\alpha _1 h),
\end{eqnarray}
where time coefficients are
\begin{eqnarray}
    \nonumber
    &&\alpha _1=\alpha _{12}= 0.079203696431196,   \\ \nonumber
    &&\alpha _2=\alpha _{11}= 0.130311410182166,   \\ \nonumber
    &&\alpha _3=\alpha _{10}= 0.222861495867608,    \\ \nonumber
    &&\alpha _4=\alpha _9=-0.366713269047426,    \\ \nonumber
    &&\alpha _5=\alpha _8=  0.324648188689706,   \\ \nonumber
    &&\alpha _6=\alpha _7= 0.109688477876750.   \nonumber
\end{eqnarray}
The long-term stability of the symplectic algorithm ensures the accuracy of numerical integration.
Building on this, the application of chaos identification tools can effectively recognize the motion states of particles.
\section{Chaos detection methods}\label{sec:five}
In the field of dynamics, commonly used chaos identification indicators include the Poincar\'{e} map  and the Lyapunov exponent.
The Poincar\'{e} map offers the advantage of visually displaying the long-term behavior and oribit structure of a system,
but it does not quantify the sensitivity or stability between orbits.
In contrast, the Lyapunov exponent offers a quantitative method for assessing the level of chaos in a system and its sensitivity to changes in initial conditions. Therefore, when studying dynamical
systems, it is often essential to combine these two methods to achieve a more comprehensive understanding and analysis.

Recently, several chaos recognition indicators\cite{Kopacek:2010yr,Cao:2024ihv,Wu:2003pe,Wu:2006rx} have been introduced and applied in the field of dynamics within general relativity.
Among them, there is a method similar to the Poincar\'{e} map called the recurrence plot, where the structure parallel to the main diagonal represents order, while points dispersed on both sides of
the main diagonal indicate chaos.
Another method for quantifying the degree of orbital chaos is the fast Lyapunov indicator,
which is an extension of the Lyapunov exponent in curved spacetime and can effectively reflect the characteristics of the system.

In this paper, we use Shannon entropy\cite{Shannon:1948,Parrondo:2015} to study its capability in recognizing chaotic and order states of particles, while also employing Poincar\'{e} map and
Lyapunov exponents to assist in validating the accuracy of the results.
Shannon entropy is an important concept in information theory and can be regard as a candidate for non-equilibrium entropy.
Its definition can be expressed using the following formula
\begin{eqnarray}
H(X) = -\sum_{i=1}^{n} p(x_i) \log_b p(x_i),
\end{eqnarray}
where $H(X)$ is the entropy of the random variable, $p(x_i)$ is the probability of the event $x_i$ occurring. $n$ is the total number of possible events. $b$ is the base of the logarithm, typically
taken as 2.

The maximum value of Shannon entropy depends on the number of possible states $n$ in the system. When each state has an equal probability $p(x_i) = \frac{1}{n}$, the entropy reaches its maximum
value, expressed as:
\begin{eqnarray}
H_{\text{max}}(X) = \log_b n.
\end{eqnarray}

For time series data, the calculation of Shannon entropy is straightforward. Generate $10^{7}$ particle coordinate
data $r$ through numerical integration of Hamilton's equations, and divide the data into 200 large intervals,
each containing $5\times10^{4}$ data points. Within each large interval, fix the range of $r$ and subdivide it into 100 small intervals.
Count the number of particles falling into each small interval to form a frequency distribution. Then, calculate the probability $p(x_i)$ for each small interval and use the Shannon entropy formula
to evaluate the uncertainty of the information.

It is worth noting that our numerical experiments always start from a stable circular orbit (ISCO) and treat it as a pure-state-like particle. When an external magnetic field acts on the particles,
periodic orbits transition to quasiperiodic and chaotic orbits, which may be regarded as a phenomenon of entropy increase. When entropy reaches its maximum value, meaning the particles have
traversed all possible states, they will transit to something-like an equilibrium state.

Next, we utilize Shannon entropy to identify the chaotic and order orbits of the three different types of black holes presented in Table. 1.
\section{Using Shannon entropy to identify chaotic and order orbits of black holes}\label{sec:six}
For each type of black hole, our calculations begin with its stable circular orbit. As the magnetic field strength increases,
chaotic behavior emerges and intensifies \cite{Kopacek:2010yr,Cao:2024ihv}. In practical applications, we gradually increase the magnetic field parameter while keeping the model parameters fixed,
testing the ability of Shannon entropy to identify the state of the orbits. We also use other chaos identification tools mentioned earlier to assist in the subsequent analysis.
\subsection{Black hole with conformal scalar hair}
We first consider a scalar hair black hole\cite{Astorino:2013sfa,Vagnozzi:2022moj}, and the metric function of the black hole solution is given by
\begin{eqnarray}
A(r)=1-\frac{2M}{r}+\frac{S}{r^{2}},
\end{eqnarray}
where $M$ is the mass of the black hole and $S$ is the hair parameter associated with the scalar field. This hair parameter $S$
emerges as an integration constant and characterizes primary hair, meaning it cannot be expressed solely as a function of the other hair parameters. The parameter $S$ can take positive or negative
values, but it must satisfy $S<M^{2}$ to ensure that the solution describes a black hole.

From Fig. 1(a), it can be seen that as charged particles move away from the black hole, their orbits gradually transition from periodic to quasi-periodic, eventually evolving into a chaotic state.
Under weaker magnetic field conditions$(b=10^{-6})$, the motion of the particles is very regular and predictable, maintaining stable orbits. However, as the magnetic field strength increases, the
range of particle motion expands, exhibiting quasi-periodicity$(b=10^{-3}~or~3\times10^{-3})$. Although the orbits still retain some structure, their complexity gradually increases. When the
magnetic field is further enhanced$(b=5\times10^{-3}~or~10^{-2})$, the system enters a chaotic state, and the motion of the particles becomes random and unpredictable, with chaotic orbits displaying
a complex distribution in phase space.

At the same time, we can draw similar conclusions from Fig. 1(b). Under weaker magnetic field conditions, charged particles undergo strict periodic motion, and during an integration time of
$10^{7}$, the calculation of Shannon entropy remains zero, indicating that the orbit is completely determined.
As the magnetic field increases, the Shannon entropy of the quasi-periodic orbit increases, indicating that the uncertainty of the orbit is also increasing, however, the Shannon entropy remains
relatively stable throughout the entire integration time. For chaotic orbits, the Shannon entropy increases further and exhibits significant fluctuations over the integration time, indicating that
the orbit is more complex and difficult to predict. However, the test particles do not reach the maximum entropy state, indicating that the system is currently in a state of increasing entropy.
Table 2 illustrates this fact more concretely. By calculating the mean, standard deviation (Std Dev), and mean absolute deviation (MAD) of the orbital entropy, we found that the Std Dev and MAD
values for chaotic orbits are significantly higher than those for order orbits. This indicates that the entropy fluctuation of chaotic orbits is stronger than that of order orbits.

The situation is similar when a charged particle is released near a black hole.
As the magnetic field strength increases, the orbit state transitions from order to chaos.
However, the structure of chaotic orbits on the section differs from that in Fig. 2(a),
it appears to be composed of interconnected periodic islands, and we found it difficult to directly distinguish chaotic orbits from order orbits in Fig. 2(b).
We then used the Lyapunov exponent, the fast Lyapunov indicator, and the Rps together to identify the orbital states.
From Fig. 2(c)-(d), it can be seen that orbits with a Lyapunov exponent tending to zero are order$(b=10^{-2}~or~2\times10^{-2})$, while orbits with a rapidly increasing FLI are
chaotic$(b=2.4\times10^{-2})$.
The Rps also provide the same results, where more complex patterns in the Rp represent chaotic
orbits(Fig. 3(a)), while a large number of diagonal structures filling the plane represent order orbits(Fig. 3(b)~and~Fig. 3(c)).

Now we can conclude that when the magnetic field parameter is $b=2.4\times10^{-2}$, the orbit is chaotic. A careful examination of the Shannon entropy for each type of orbit also leads to the same
conclusion. From Fig. 4(a)-(c), it can be seen that the Shannon entropy of the three types of orbits exhibits fluctuations. The time series data of chaotic orbits is irregular, while the time series
data of order orbits shows a certain periodicity. This observation becomes apparent after performing a fast Fourier transform on the time series data. The time series of entropy for order orbits
exhibits a limited number of dominant modes, while the entropy for chaotic orbits displays multiple dominant modes. Similarly, we can directly conclude from Table 3 that the entropy fluctuation of
chaotic orbits is stronger than that of order orbits, and orbit 5 is a chaotic orbit.
\subsection{Brane-world black hole}
We now examine one of the most compelling models in high-energy physics: the Randall-Sundrum II (RSII) brane-world model.
This model has been explored across a broad range of scales and assessed using a diverse array of experimental probes\cite{Vagnozzi:2022moj,Antoniadis:1998ig,Johannsen:2008aa}.
The metric function for a spherically symmetric black hole is expressed as follows
\begin{eqnarray}
A(r)=1-\frac{2M}{r}+\frac{4M^{2}q}{r^{2}},
\end{eqnarray}
where $q$ characterizes a specific hair.
From Fig. 5(a)-(b), it can be seen that the trajectories marked in red$(b=3\times10^{-3})$ and
magenta-marked $(b=1.9741\times10^{-3})$ trajectories are both chaotic,
corresponding to the random scatter points in the Poincar\'{e} section and showing larger fluctuations in entropy values.
From time series data, it can be observed that the entropy of order trajectories exhibits a certain periodicity,
while the entropy of chaotic trajectories is very irregular.
Through Fourier transform, it can be more clearly observed that the time series data of the entropy of order trajectories exhibits five dominant modes, whereas the time series data of the entropy of
chaotic trajectories shows multiple modes, indicating the complexity of the system. The specific entropy fluctuations can be referenced from Table 4, and the results show that the entropy
fluctuations of chaotic orbits are still stronger than those of order orbits.

\subsection{Einstein-{\ae}ther gravity}
We further test the black holes within the framework of Einstein-{\ae}ther theory. Due to the presence of the {\ae}ther field, Lorentz symmetry is locally broken. In this theory, the second exact
black hole solution is applicable for $c_{14}\neq0$
and $c_{123}=0$\cite{Vagnozzi:2022moj,Wang:2022yvi}. The metric function of this spacetime is given by
\begin{eqnarray}
A(r)=1-\frac{2M}{r}+\frac{M^{2}(c_{14}-2c_{13})}{2r^{2}(1-c_{13})},
\end{eqnarray}
where $c_{13}$ and $c_{14}$ obviously characterize universal hairs.

From Fig. 6(a), it can be observed that as the magnetic field increases,
the periodic island chain on the Poincar\'{e} section gradually merges into a single closed curve,
followed by the emergence of chaos.
However, from Fig. 6(b), we observe that as the magnetic field increases, the Shannon entropy rises, indicating an increase in the uncertainty of the trajectories, and there are no significant
fluctuations observed in the chaotic trajectories.
Through the fast Fourier transform, it can be observed from Fig. 6(c)-(f) that the time series data of the Shannon entropy for the periodic island chain trajectories exhibits seven dominant modes,
whereas the chaotic trajectories display multiple dominant modes. From Table 5, we observe that due to the periodic distortion of the periodic island chain, the average entropy of Orbit 2 is
unexpectedly higher than that of the chaotic orbit. This indicates that a high entropy state does not necessarily represent a chaotic orbit. However, through the analysis of Std Dev and MAD, we can
see that the fluctuations of the chaotic orbit is still stronger than that of the order orbits.

\subsection{Kerr black hole}
In astronomy, the Kerr solution is widely applied, primarily because it precisely describes the spacetime structure of rotating celestial bodies, offering a depiction that is closer to the physical
reality of the universe. Ultimately, we  investigate the fluctuations of orbital entropy in the Kerr spacetime background. In Fig. 7(a), we tested five trajectories, with the initial particle
release position gradually increasing, causing the orbital state to transition from order to chaotic. By observing the Poincaré map, we can easily draw this conclusion. In Fig. 7(b), we obtained the
same conclusion. The chaotic orbit exhibits significant entropy fluctuations, while the order orbit shows smaller entropy fluctuations. According to Table 6, the fluctuation value of the chaotic
orbit is much greater than that of the order orbit. This suggests that the existence of orbital entropy fluctuations is possible in axially symmetric spacetimes.

\section{Conclusions}\label{sec:seven}
We propose that Shannon entropy can serve as a sensitive probe for detecting chaotic states of particles in curved spacetimes.

Under the condition of source-free Maxwell equations in vacuum, we derive the solutions for the Wald potential of three types of Reissner-Nordstr\"{o}m-like black holes. We also calculated the
Shannon entropy for particle motion both near and far from the black hole.
We found that for stable circular orbits, which are periodic orbits, the Shannon entropy is zero, indicating that the system is completely determined.
For orbits which states from regular to chaotic, the entropy values increase, representing an increase in orbital uncertainty.
The Shannon entropy of chaotic orbits exhibits significant fluctuations over time, while order orbits show minimal fluctuations, providing an intuitive method to distinguish them.

For the special case of chaotic orbits with small entropy fluctuations, we performed a fast Fourier transform on their time series.
The results indicate that the entropy time series of order orbits has fewer dominant modes, while chaotic orbits exhibit multiple dominant modes.
In these models, the entropy fluctuations (Std Dev) of the chaotic orbit are always stronger than those of the order orbit, and a higher entropy state does not necessarily represent a chaotic orbit.
Our study suggests that Shannon entropy may serve as an adiabatic invariant in system of relativistic particles. As a comparison, the Gibbs entropy in microcanonical ensemble is an adiabatic
invariant\cite{Park:2022jkps}.

We show that the Shannon entropy can be used to calculate the entropy of specific orbits. When an external magnetic field perturbs spacetime, the orbit becomes quasi-periodic and then go into
chaotic state,  accompanied by an increase of entropy in our definition.

\section*{Acknowledgments}
We would like to thank Dr. Shiyang Hu  for numerous helpful discussions.
This work is supported by the National Natural Science
Foundation of China (NSFC) under Grant nos. 12275106, 12235019
and Shandong Provincial Natural Science Foundation under grant No. ZR2024QA032.

\begin{table*}[htbp]
\centering \caption{The Wald potential of non-vacuum spherically symmetric black holes and the constraints on black hole model parameters derived from black hole shadows.\label{tab1}}
\scriptsize
\begin{tabular}{ccccccc}
\hline
   &Black hole&   $f_{2}$ & $A_{\mu}$ &Constraints&\\
\hline
&BH with conformal scalar hair & $S$ & $\frac{B}{2}  \left(r^2-S\right) \sin ^2(\theta )$ & $ 0\leq S\leq0.9$&\\
\hline
&Brane-world BH&$4M^{2}q$&$\frac{B}{2} \left(r^2-4M^{2}q\right) \sin ^2(\theta )$&$0\leq q\leq 0.2$&\\
\hline
&Einstein-{\ae}ther BH&$\frac{M^{2}(c_{14}-2c_{13})}{2(1-c_{13})}$&$\frac{B}{2} \left(r^2-\frac{M^{2}(c_{14}-2c_{13})}{2(1-c_{13})}\right) \sin ^2(\theta )$&$c_{13}<1,~0\leq c_{14}\leq2$&\\
\hline
\end{tabular}
\end{table*}

\begin{figure*}[htbp]
\center{
\includegraphics[scale=0.2]{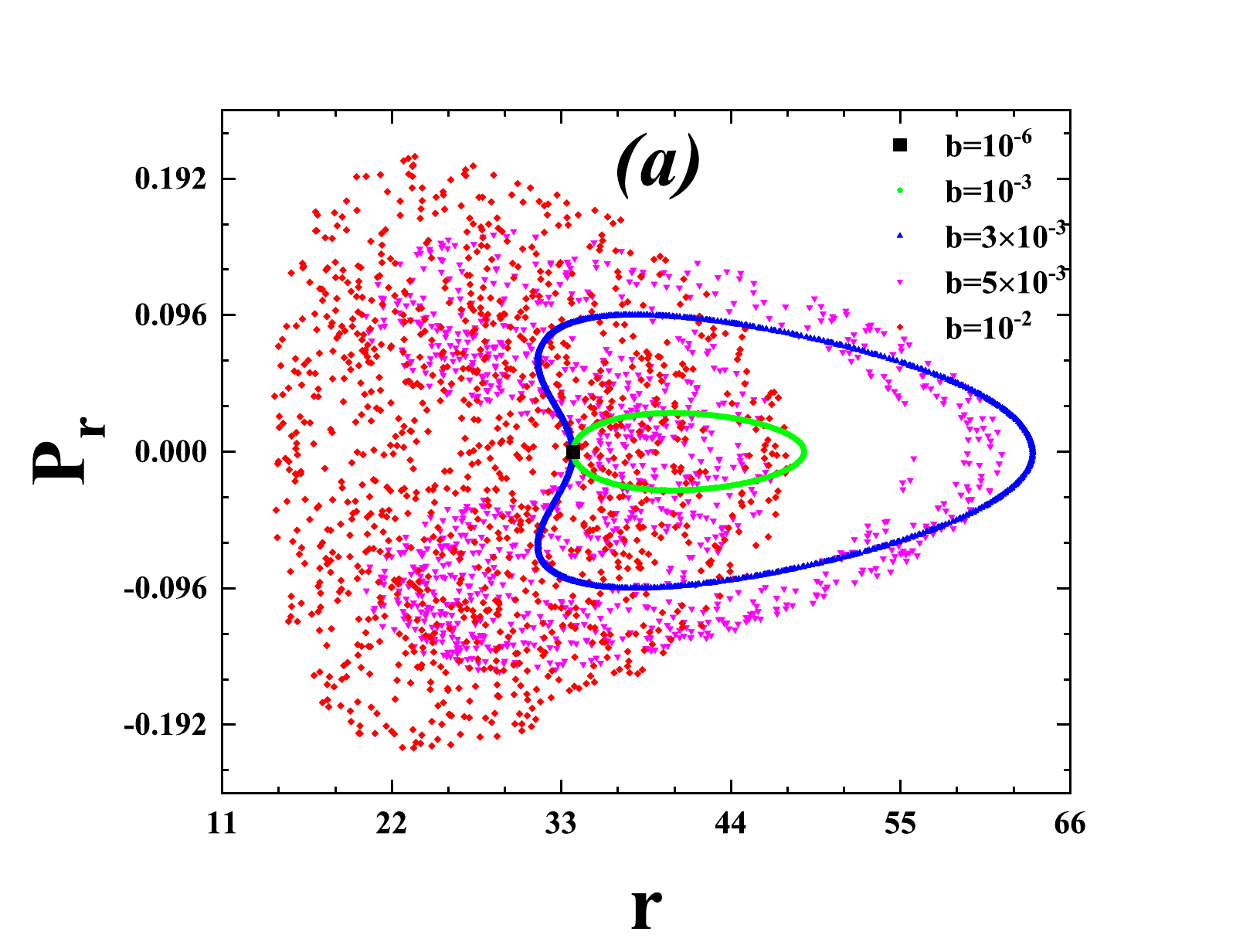}
\includegraphics[scale=0.2]{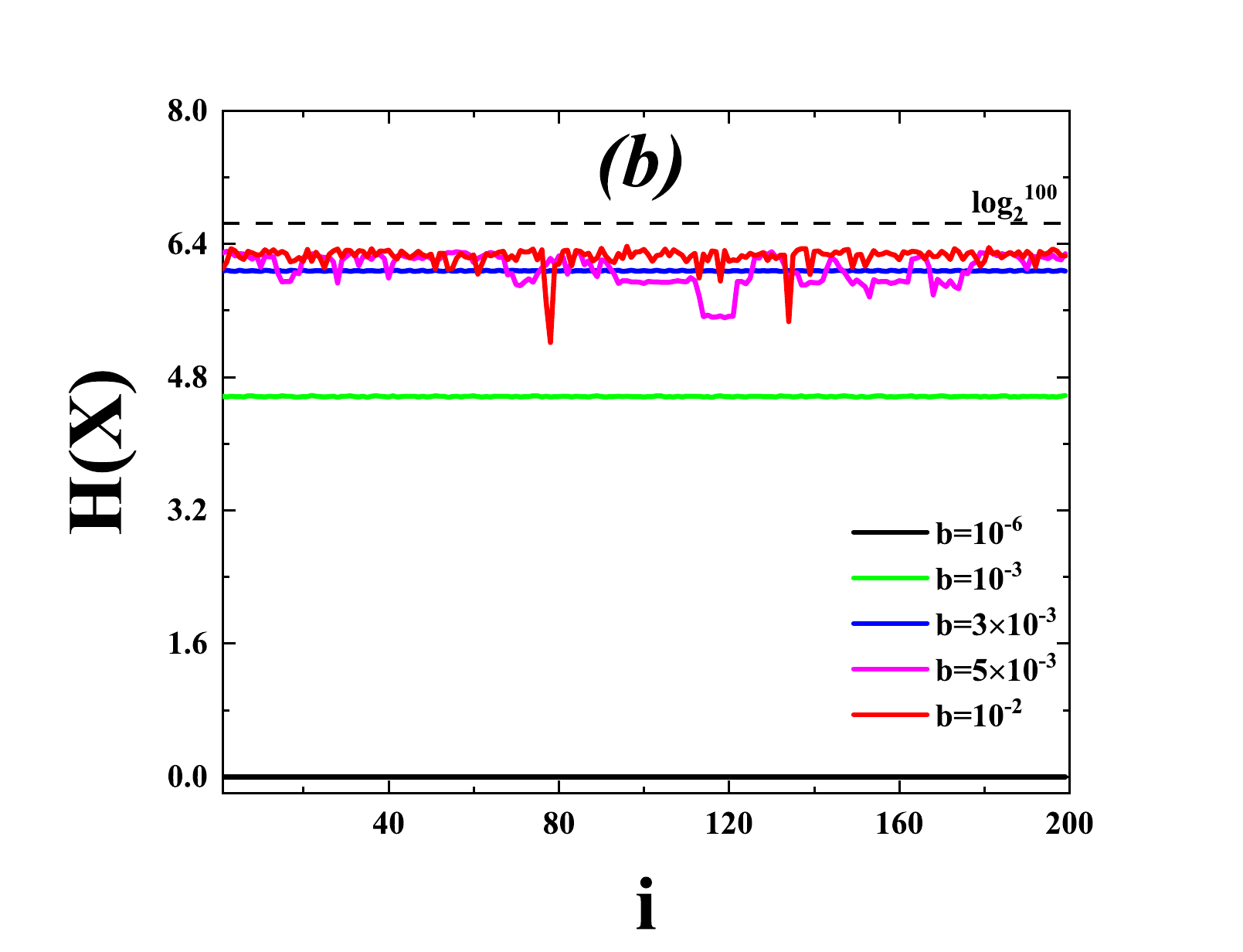}
\caption{The Poincar\'{e} map and Shannon entropy of charged particles moving around a black hole in an external magnetic field.
The initial positions where the particles are released at $S=0.9$, $r=33.757$, $\theta=\frac{\pi}{2}$, $E=0.985$, and $L=6$.
(a): Poincar\'{e} map under different magnetic field conditions.
(b): Shannon entropy under different magnetic field conditions.}
 \label{Fig1}}
\end{figure*}

\begin{table}[htbp]
\centering
\caption{Fluctuation Analysis of Orbital Entropy in Figure 1}
\begin{tabular}{lrrrrl}
\toprule
\textbf{Orbital ID}&$b$&\textbf{Mean}&\textbf{Std Dev}&\textbf{MAD}\\
\midrule
Orbit 1  &$10^{-6}$& 0&0 &0\\
Orbit 2  &$10^{-3}$& 4.57 &0.005&0.0039\\
Orbit 3  &$3\times10^{-3}$&6.08& 0.004&0.0037\\
Orbit 4  &$5\times10^{-3}$&6.10& 0.187&0.158\\
Orbit 5  &$10^{-2}$& 6.25&0.121&0.065\\
\bottomrule
\end{tabular}
\end{table}

\begin{figure*}[htbp]
\center{
\includegraphics[scale=0.2]{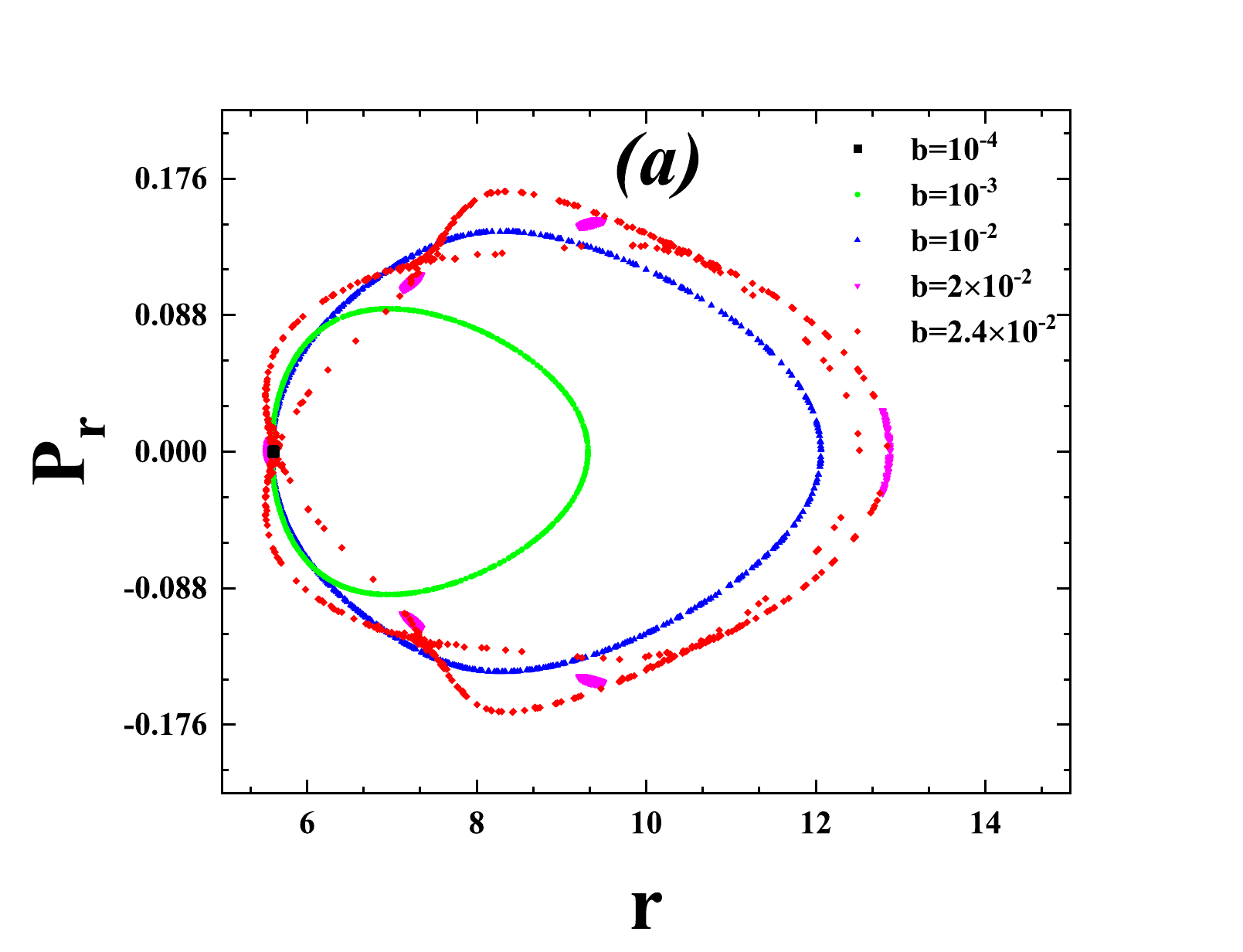}
\includegraphics[scale=0.2]{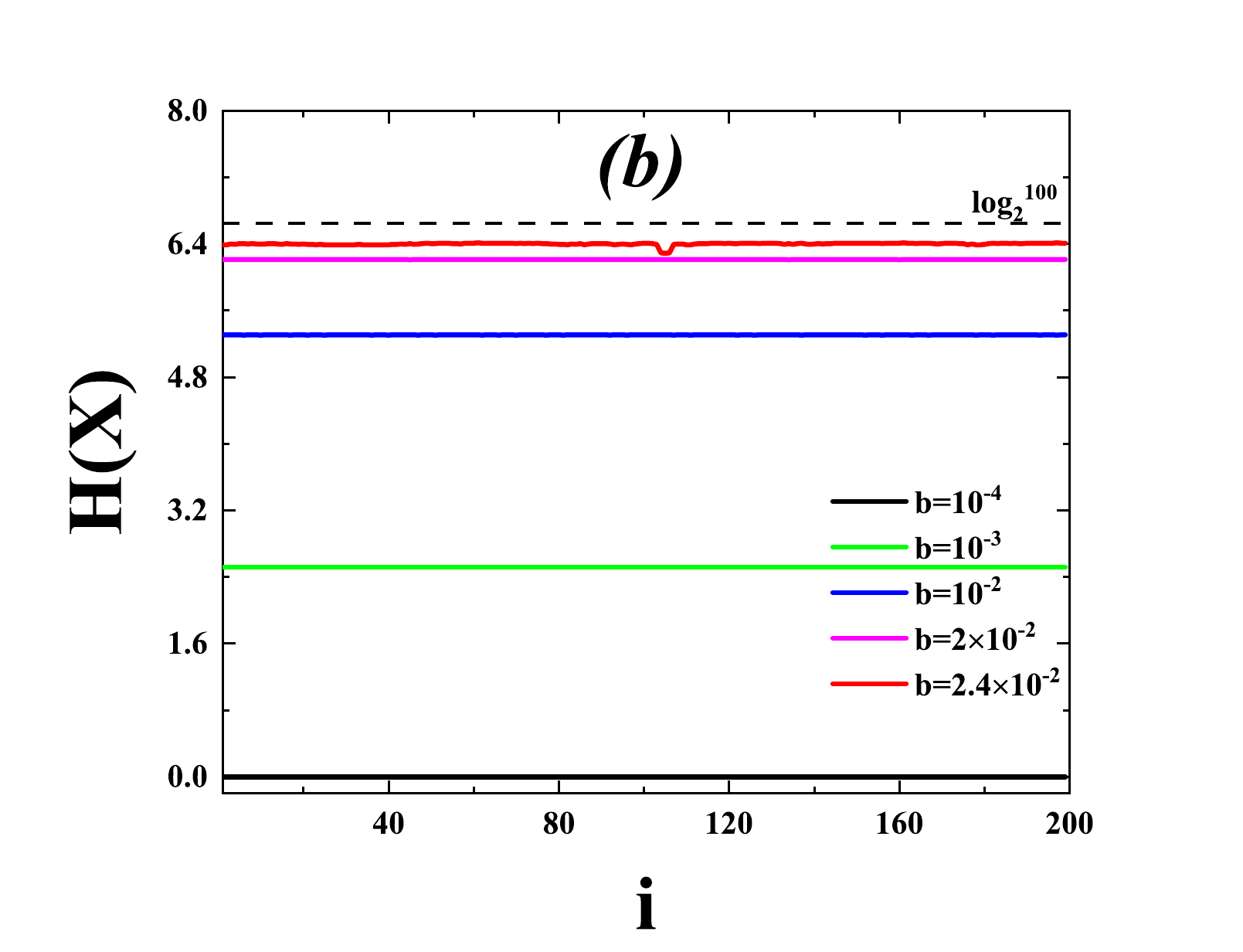}
\includegraphics[scale=0.2]{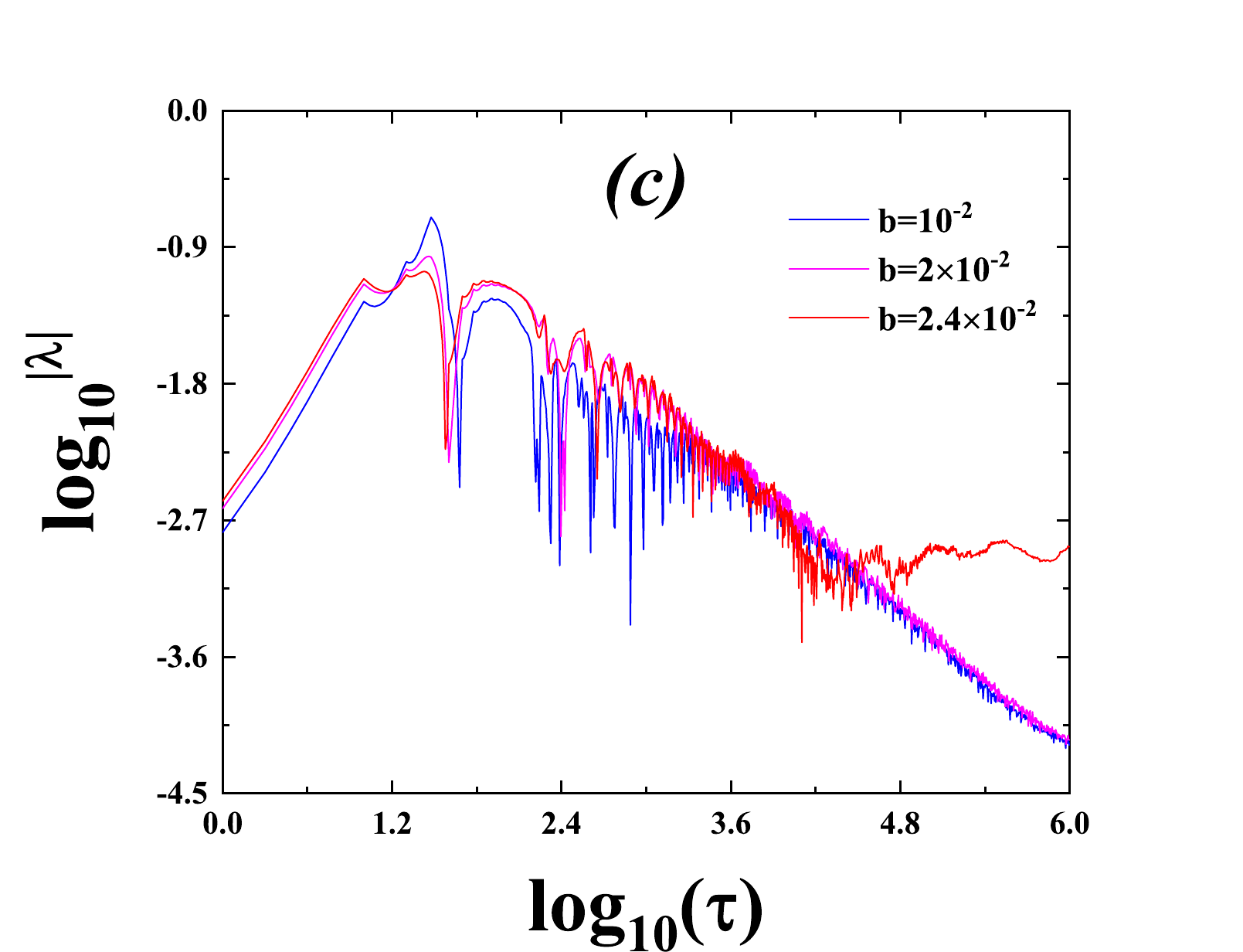}
\includegraphics[scale=0.2]{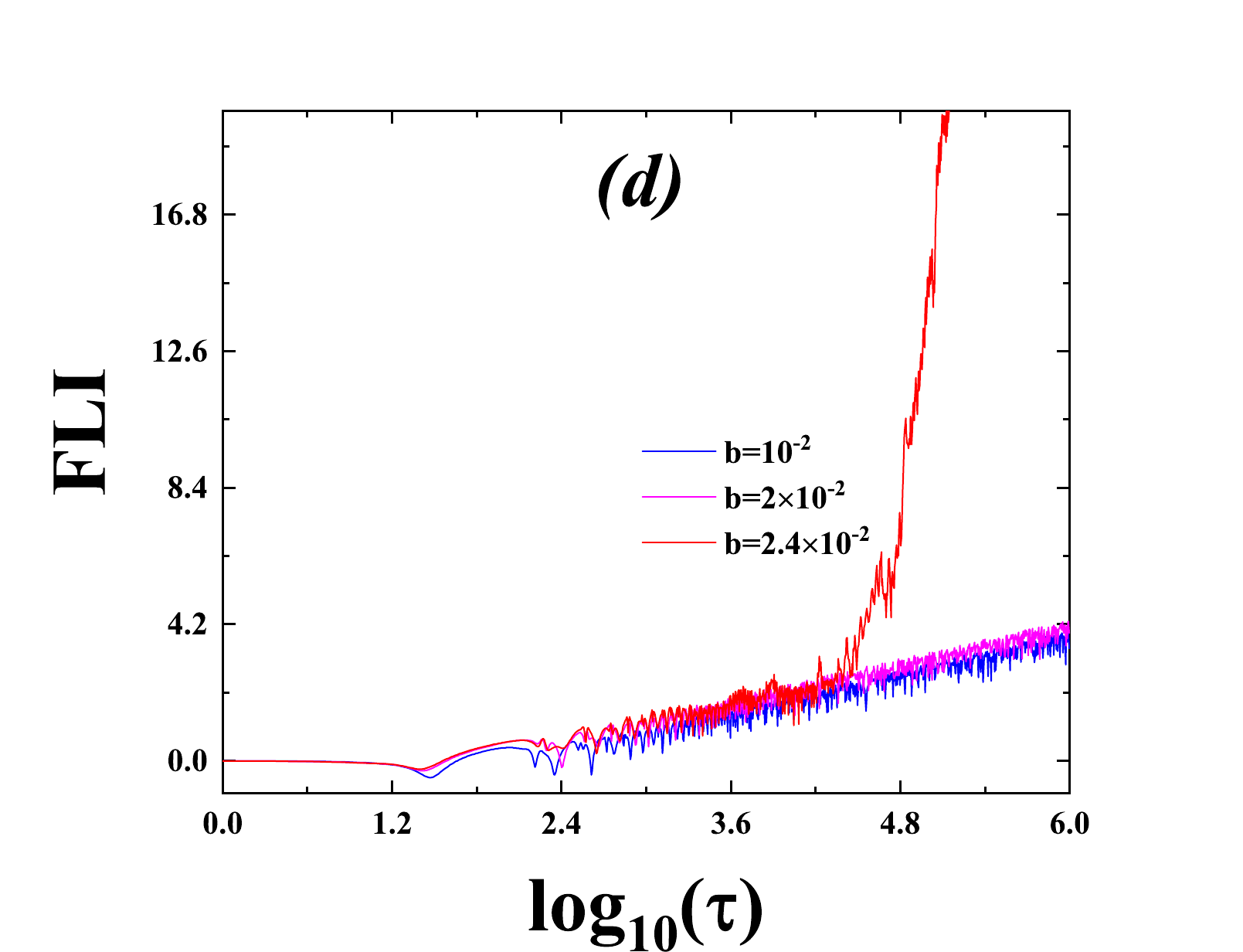}
\caption{The four chaotic indicators of charged particles moving around a black hole in an external magnetic field.
The initial positions where the particles are released at $S=0.9$, $r=5.584$, $\theta=\frac{\pi}{2}$, $E=0.929$, and $L=3$.
(a): Poincar\'{e} map under various magnetic field conditions.
(b): Shannon entropy under various magnetic field conditions.
(c): Lyapunov exponents of order and chaotic orbits.
(d): Fast Lyapunov indicators of order and chaotic orbits.}
 \label{Fig2}}
\end{figure*}

\begin{figure*}[htbp]
\center{
\includegraphics[scale=0.25]{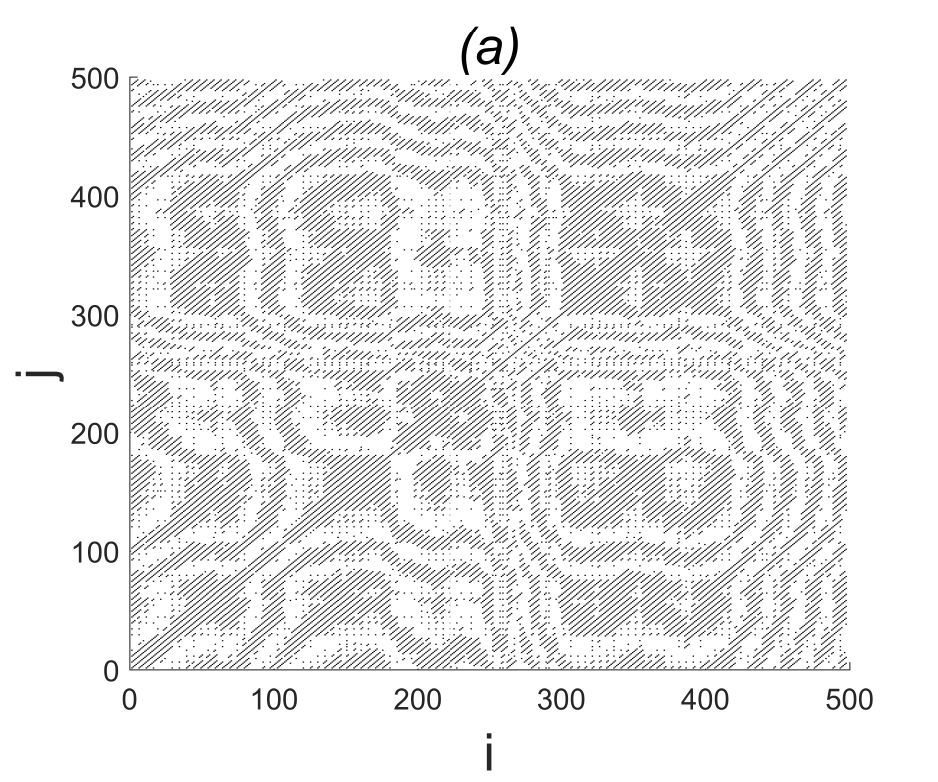}
\includegraphics[scale=0.25]{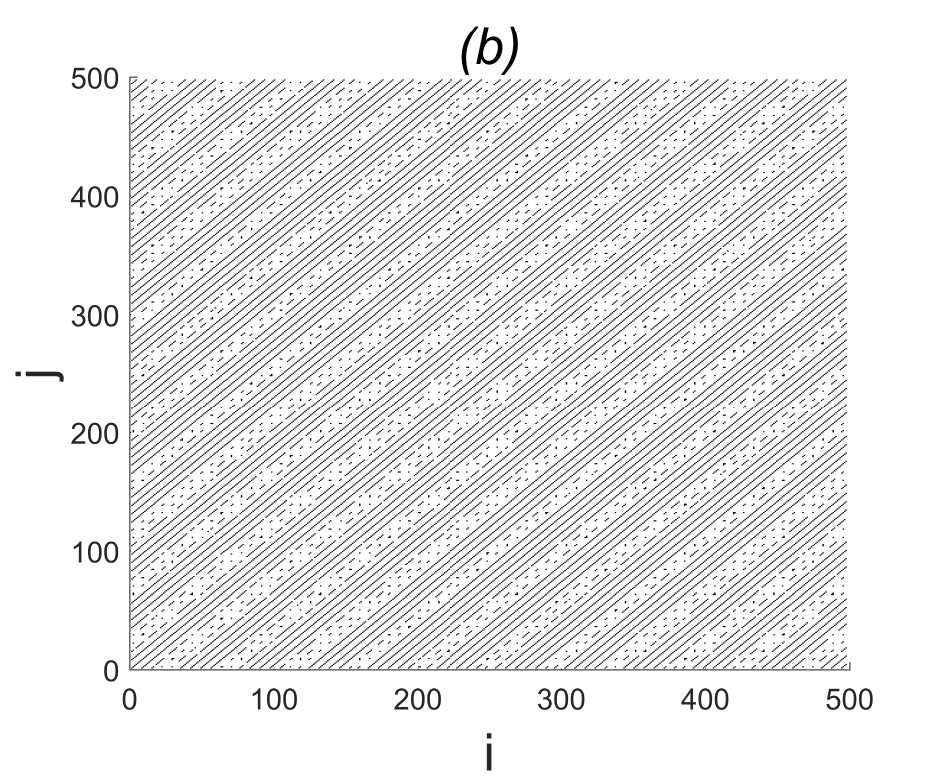}
\includegraphics[scale=0.25]{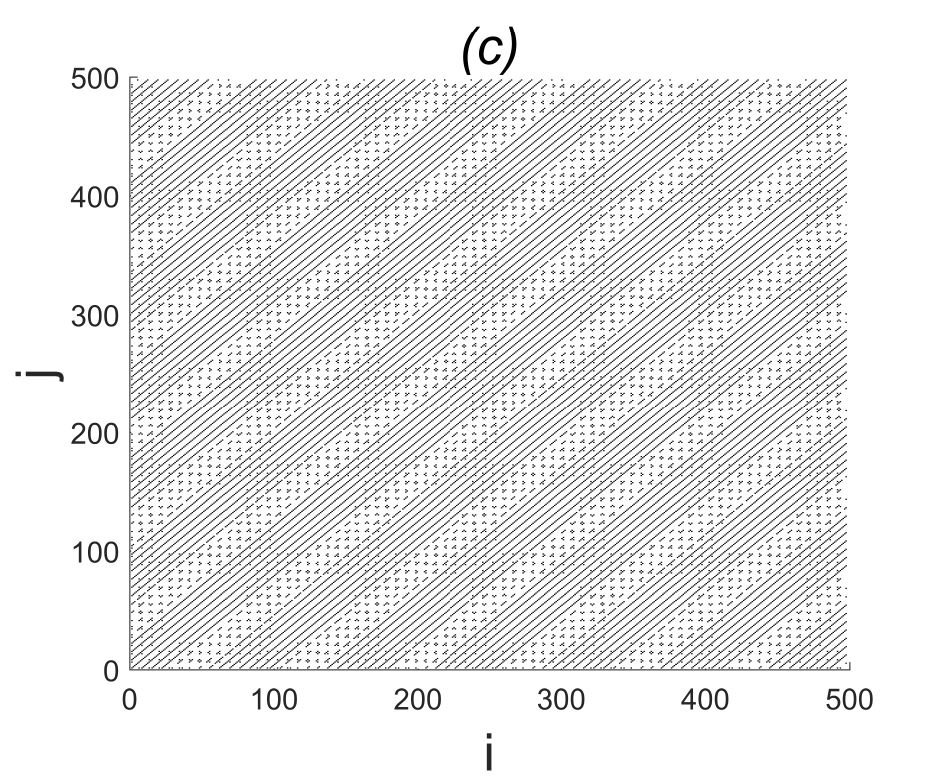}
\caption{Recurrence plots of order and chaotic orbits. From left to right:
 $b=2.4\times10^{-2}$ ,$2\times10^{-2}$ and $10^{-2}$.}
 \label{Fig1}}
\end{figure*}

\begin{figure*}[htbp]
\center{
\includegraphics[scale=0.2]{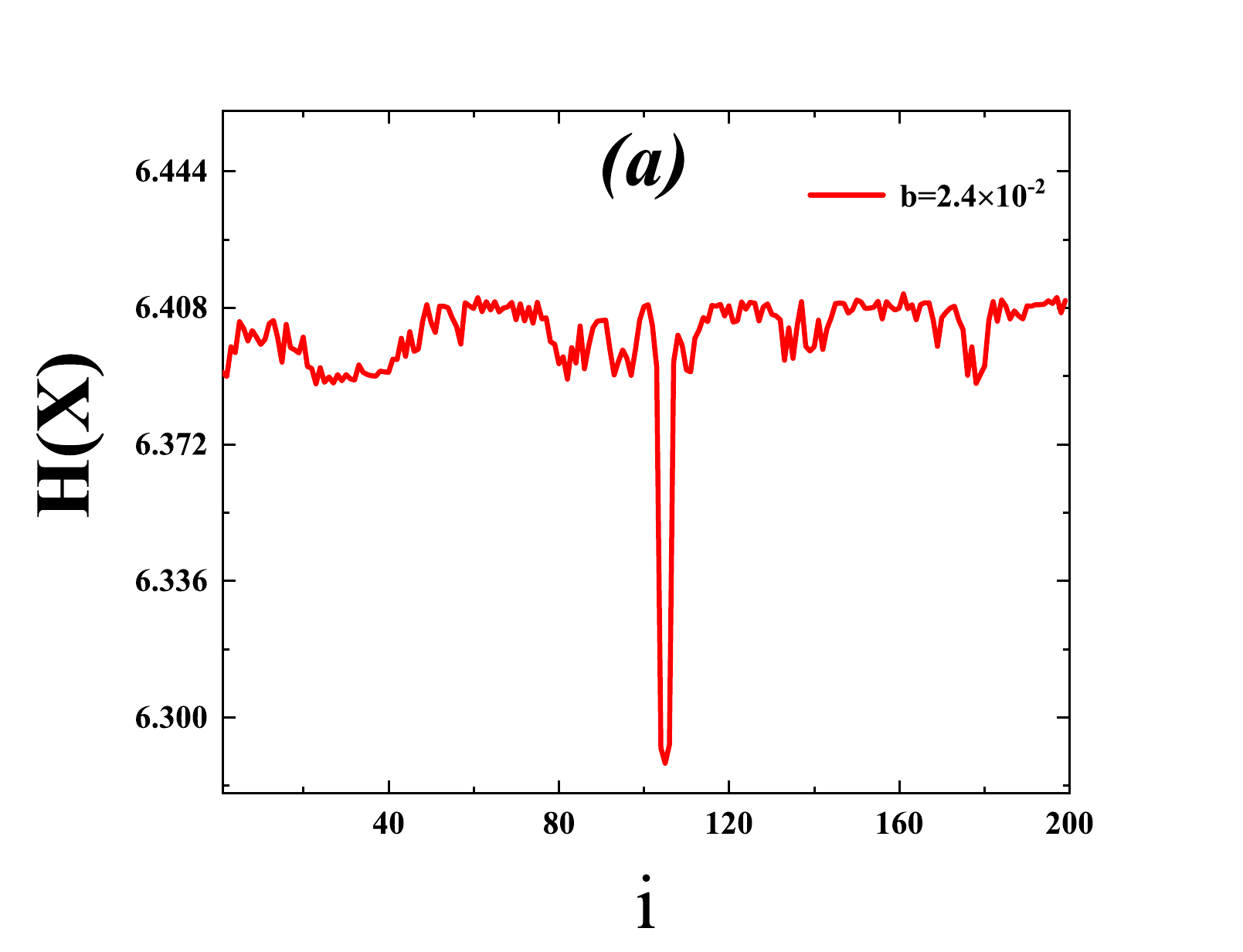}
\includegraphics[scale=0.2]{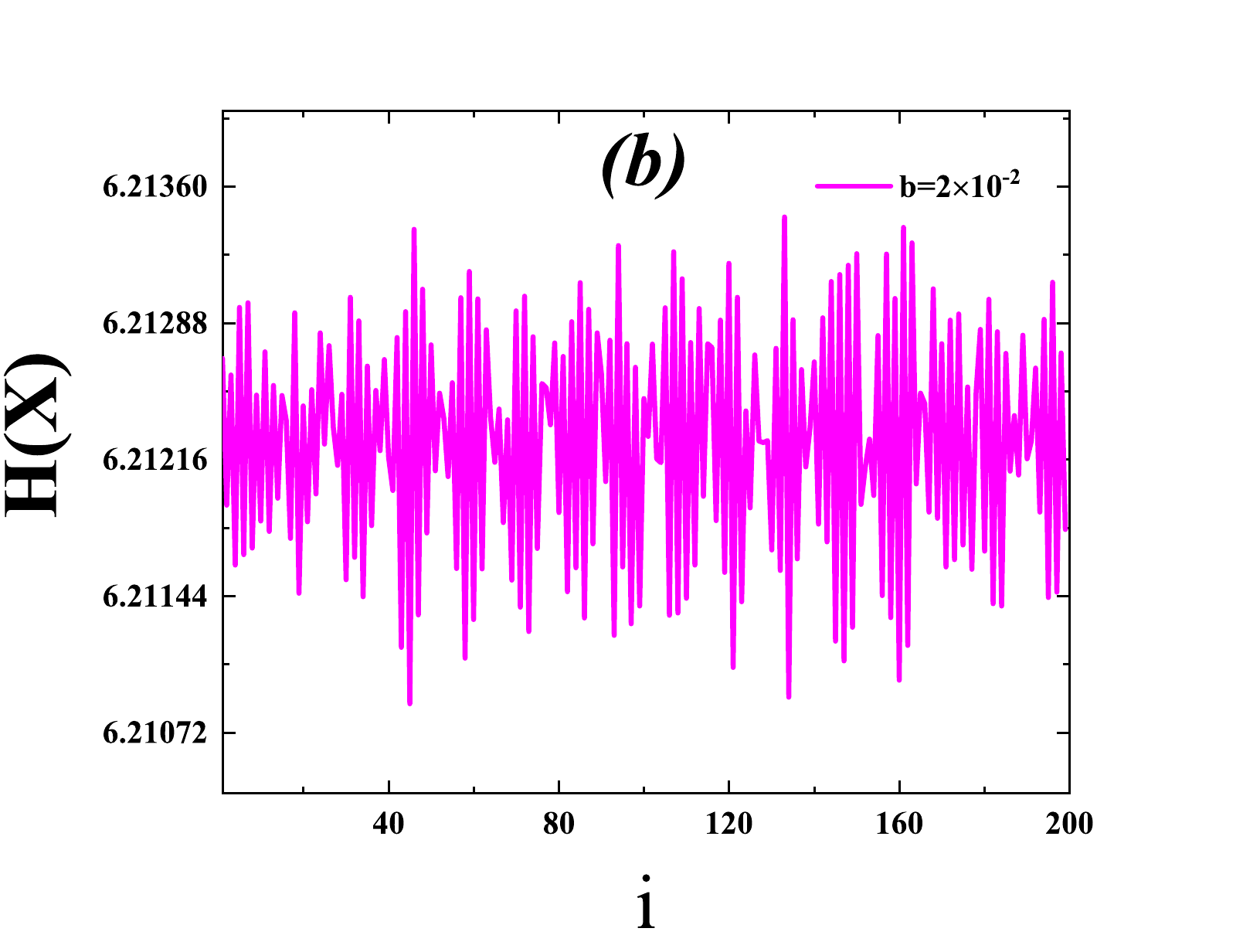}
\includegraphics[scale=0.2]{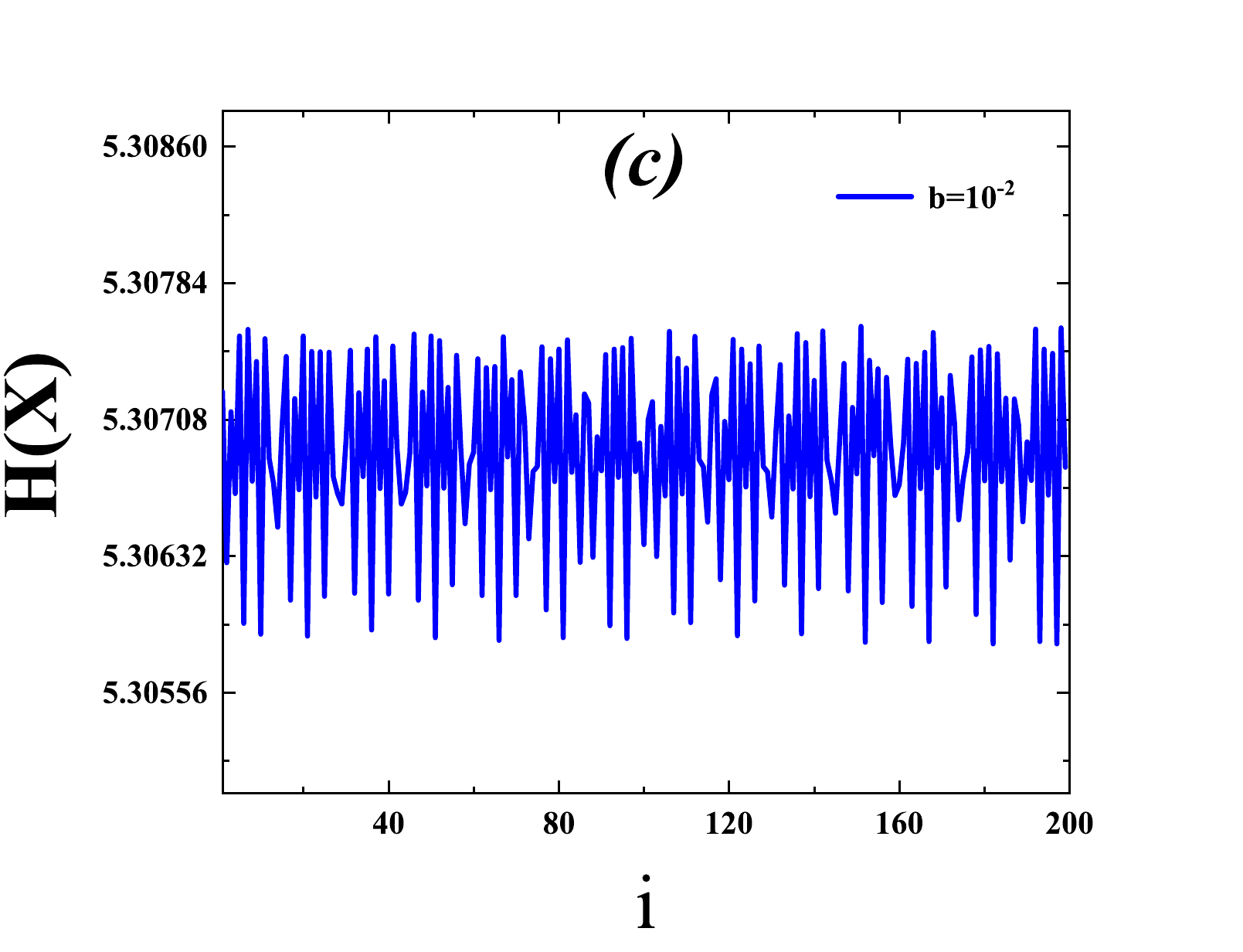}
\includegraphics[scale=0.3]{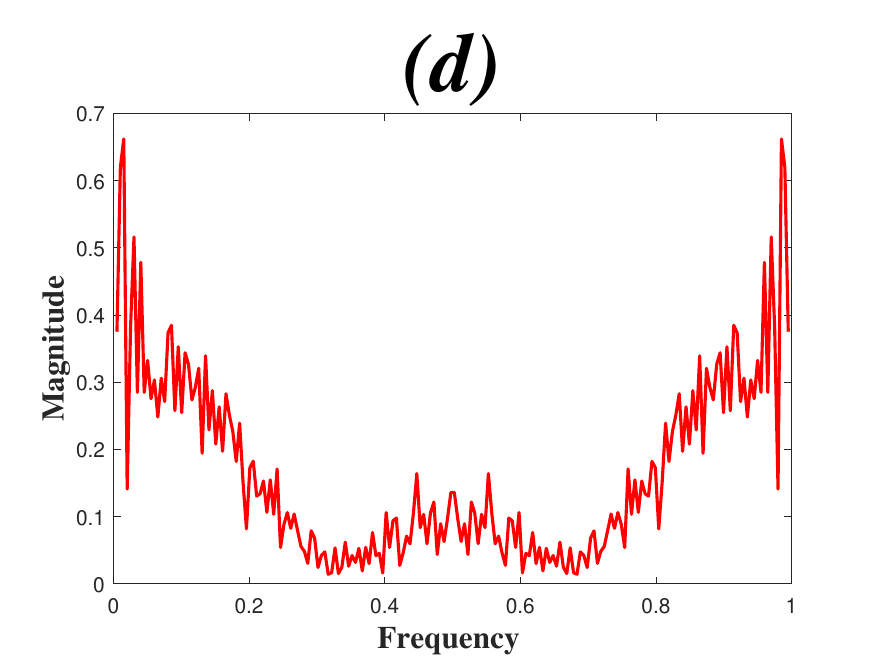}
\includegraphics[scale=0.3]{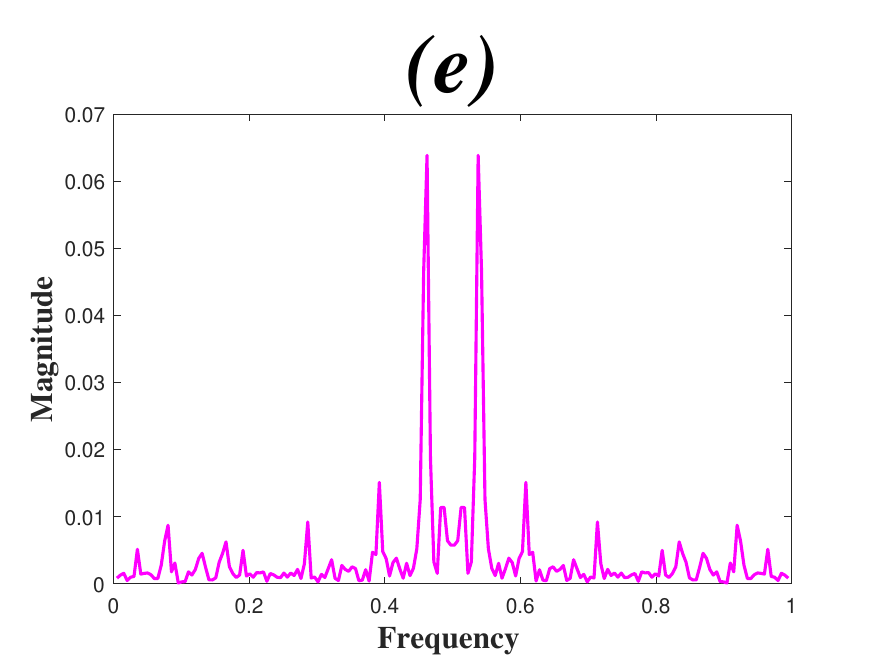}
\includegraphics[scale=0.3]{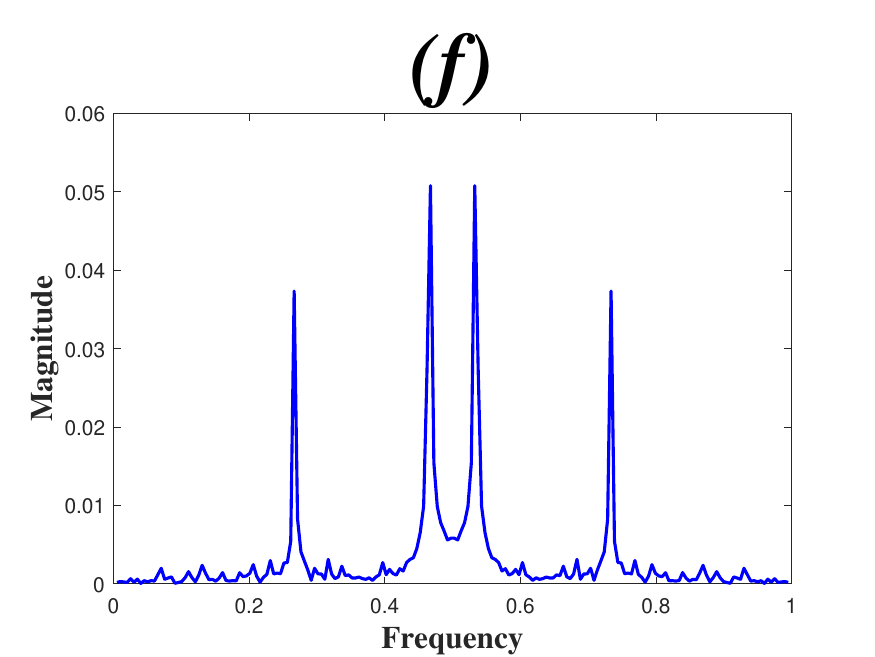}
\caption{Shannon entropy of three types of orbits and their fast Fourier analysis in Fig. 2(b),
with colors corresponding to those in the same figure.}
 \label{Fig3}}
\end{figure*}

\begin{table}[htbp]
\centering
\caption{Fluctuation Analysis of Orbital Entropy in Figure 2}
\begin{tabular}{lrrrrl}
\toprule
\textbf{Orbital ID}&$b$&\textbf{Mean}&\textbf{Std Dev}&\textbf{MAD}\\
\midrule
Orbit 1  &$10^{-4}$& 0&0 &0\\
Orbit 2  &$10^{-3}$&  2.52& 0.0002& 0.0002\\
Orbit 3  &$10^{-2}$&5.31& 0.0005& 0.0004\\
Orbit 4  &$2\times10^{-2}$&6.21&0.0006&0.0006\\
Orbit 5  &$2.4\times10^{-2}$&6.40& 0.015&0.0079\\
\bottomrule
\end{tabular}
\end{table}

\begin{figure*}[htbp]
\center{
\includegraphics[scale=0.2]{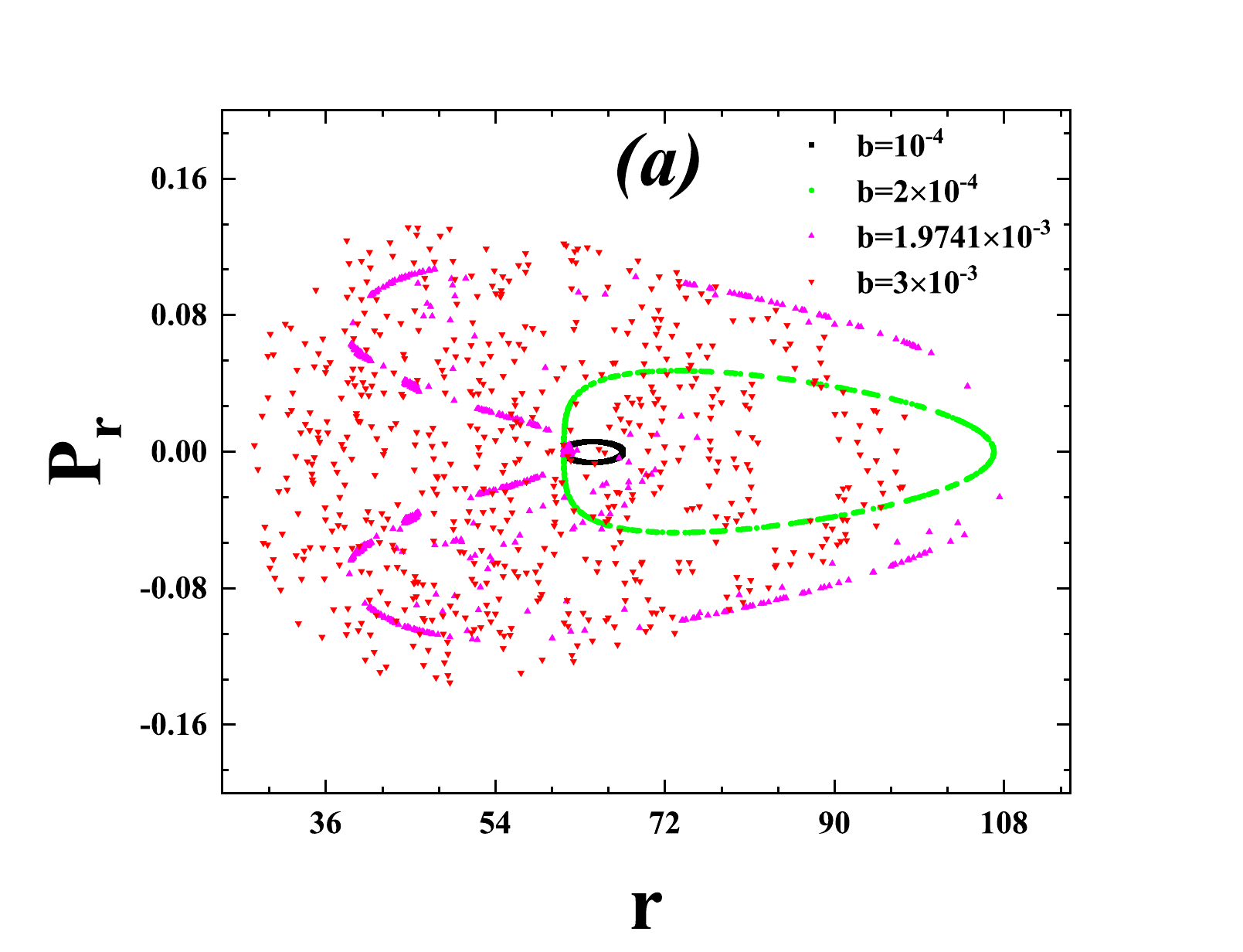}
\includegraphics[scale=0.2]{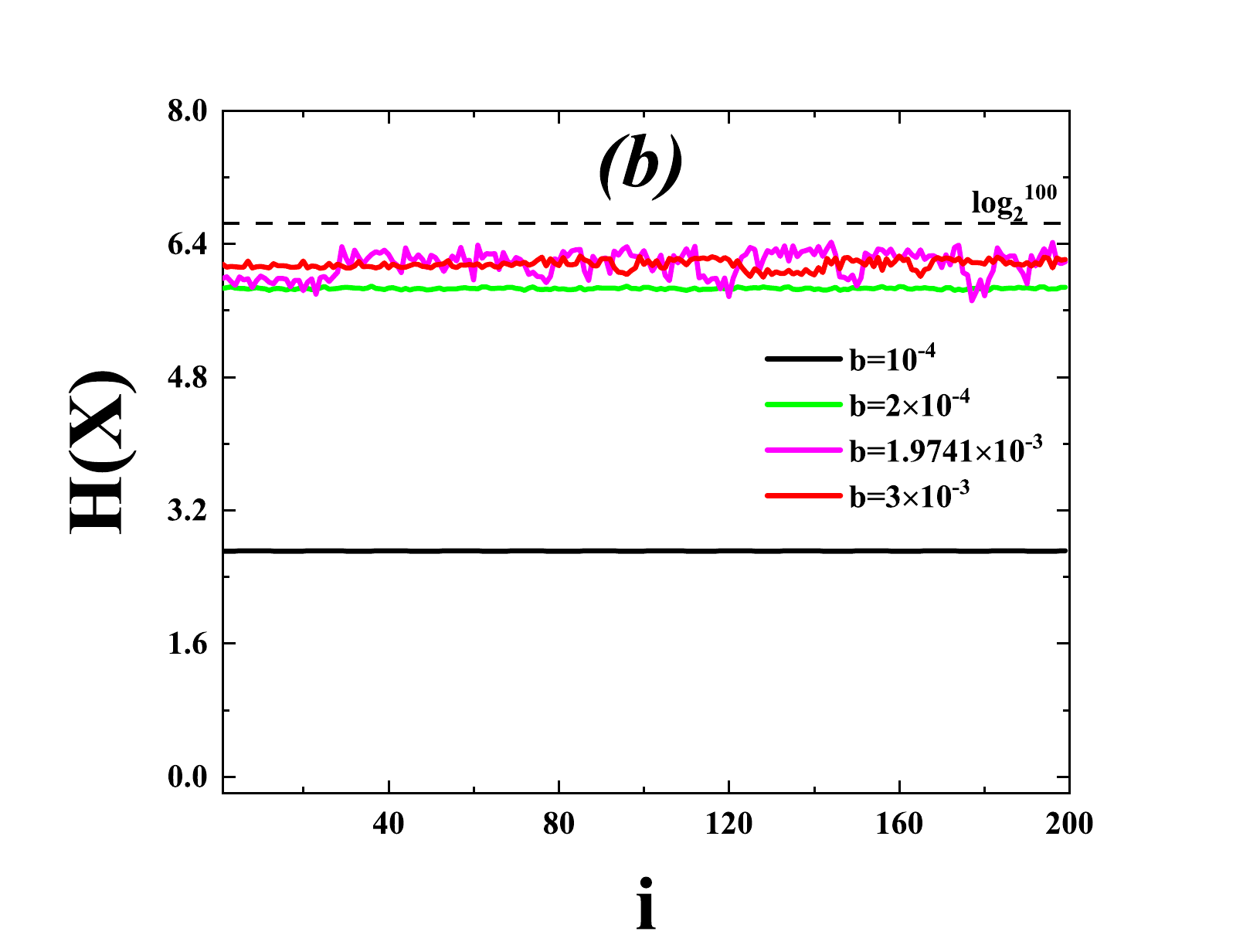}
\includegraphics[scale=0.2]{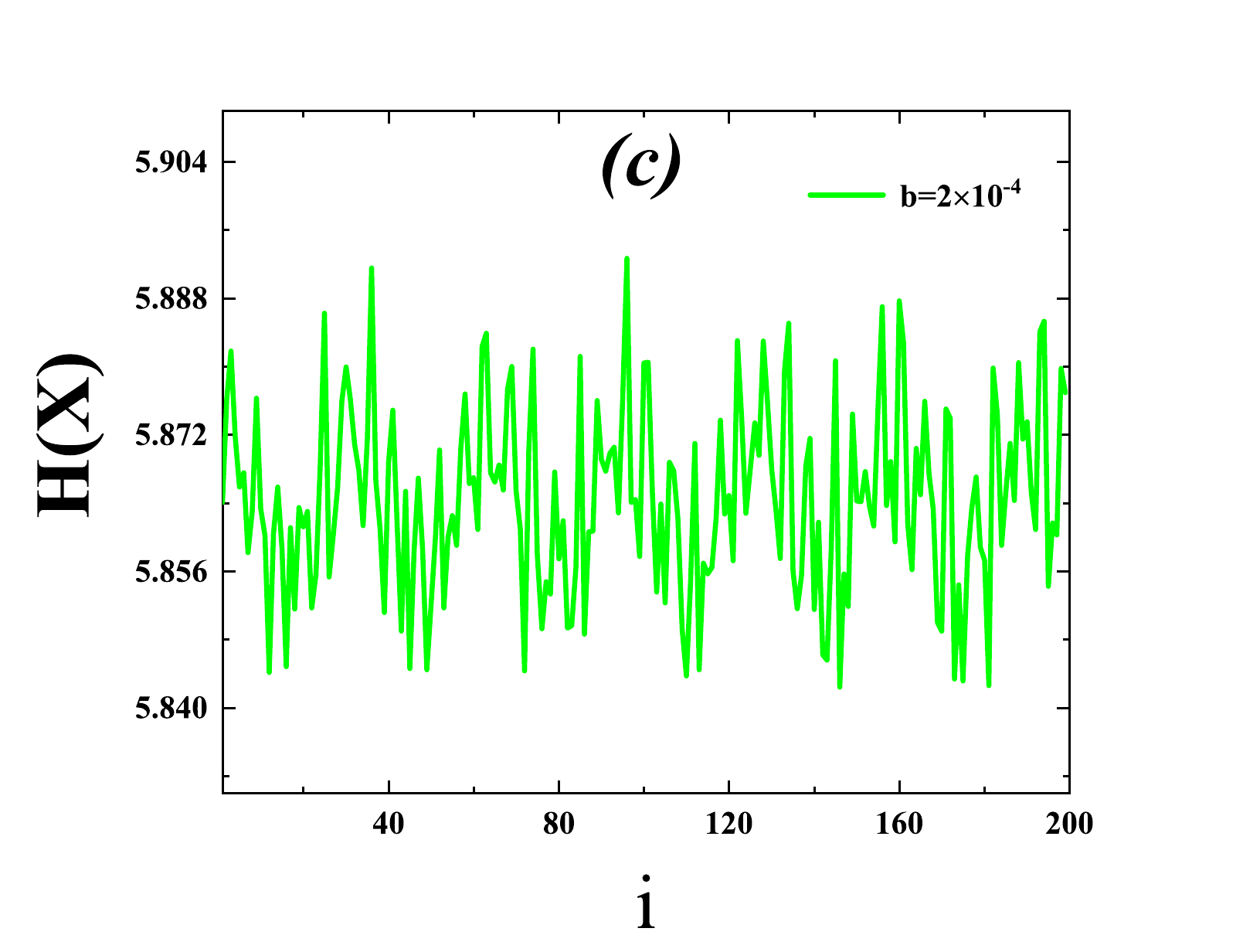}
\includegraphics[scale=0.2]{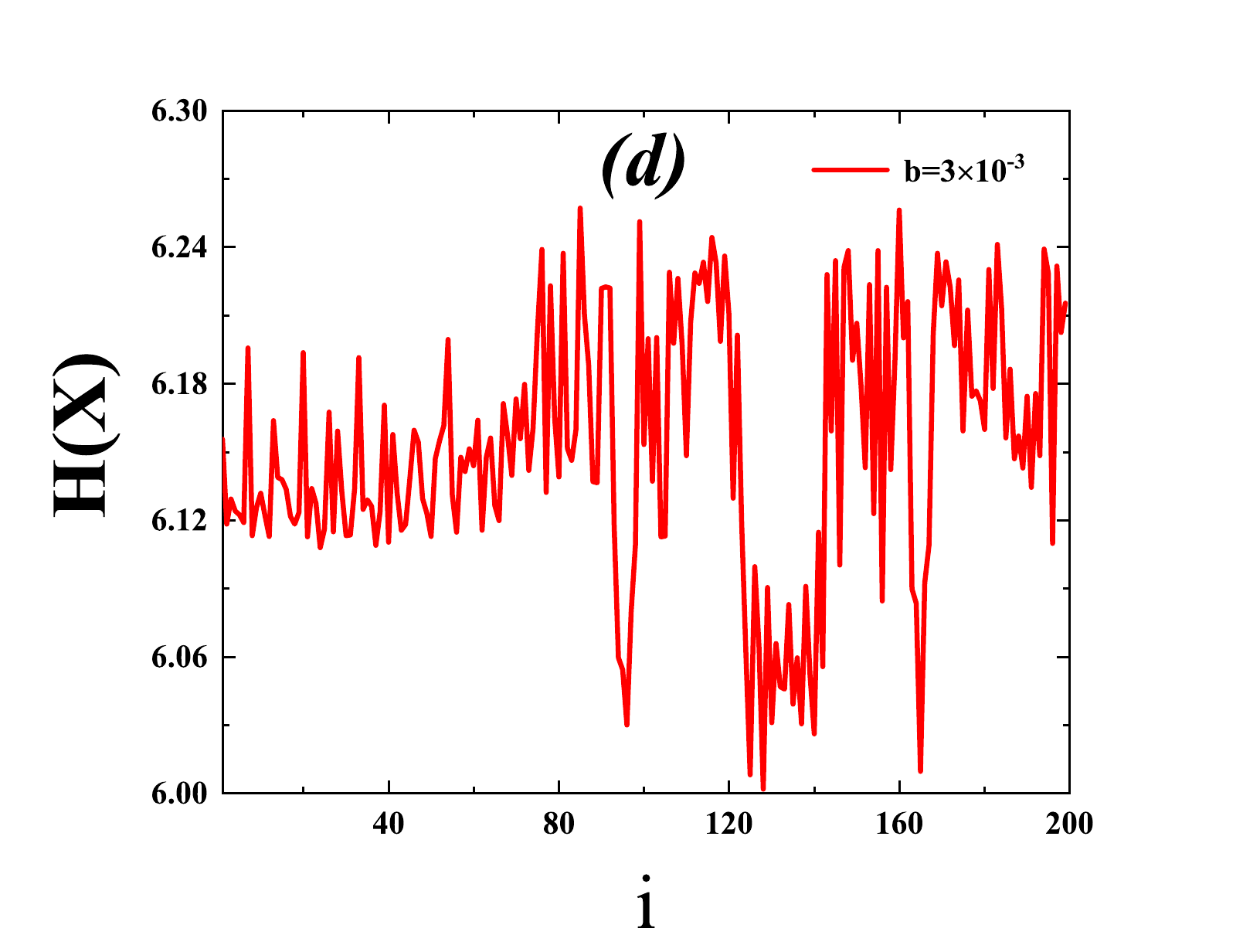}
\includegraphics[scale=0.3]{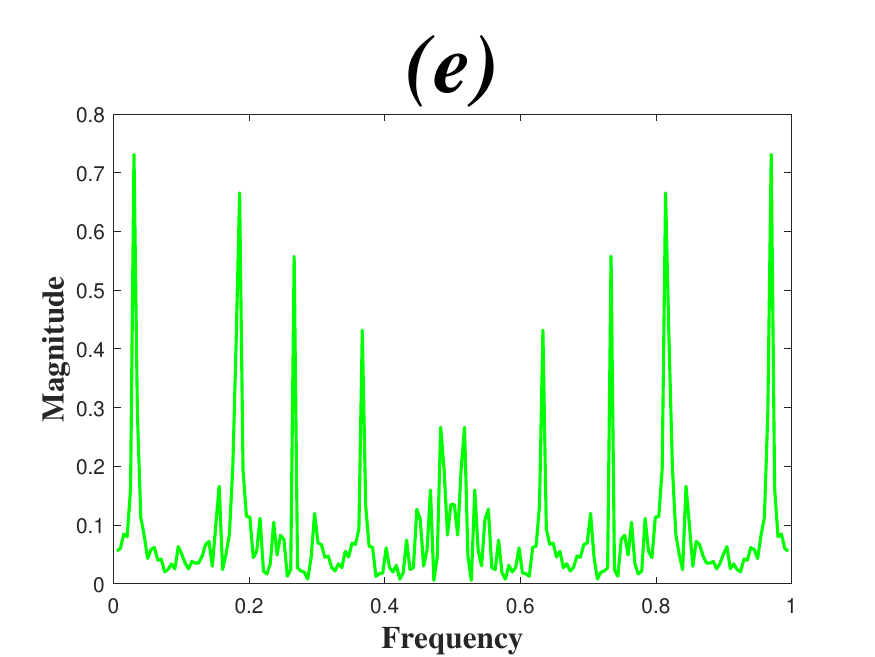}
\includegraphics[scale=0.3]{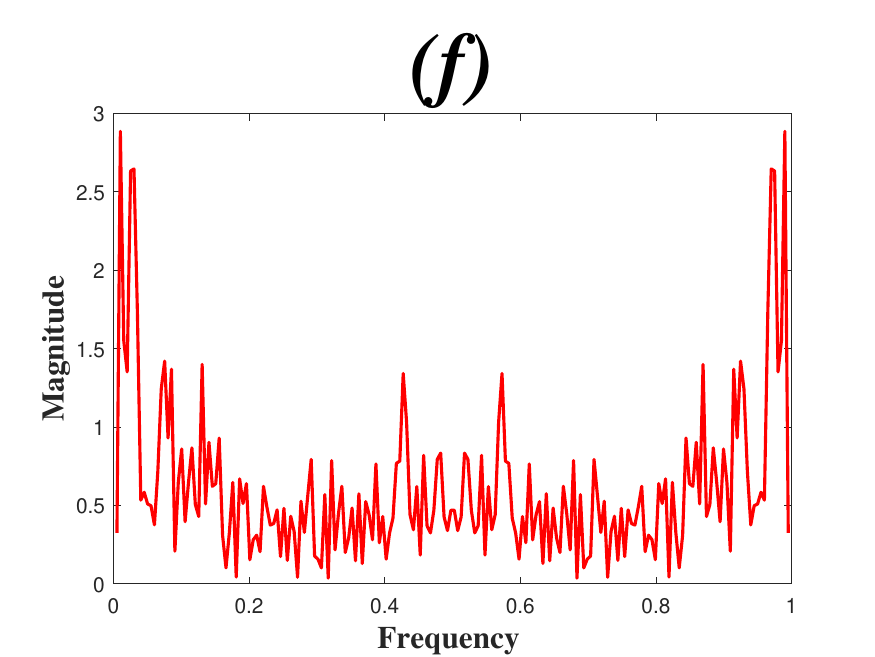}
\caption{The Poincar\'{e} map and Shannon entropy of charged particles moving around a black hole in an external magnetic field.
The initial position of the particle release is $Q=0.1$, $r=61.281$, $\theta=\frac{\pi}{2}$, $E=0.992$, and $L=8$.
(a): The Poincar\'{e} section of charged particles as the magnetic field strengthens.
(b): The Shannon entropy of charged particles as the magnetic field strengthens.
(c)-(f): Shannon entropy of two types of orbits and their fast Fourier analysis in Fig. 5(b), with colors corresponding to those in the same figure.}
 \label{Fig4}}
\end{figure*}

\begin{table}[htbp]
\centering
\caption{Fluctuation Analysis of Orbital Entropy in Figure 5}
\begin{tabular}{lrrrrl}
\toprule
\textbf{Orbital ID}&$b$&\textbf{Mean}&\textbf{Std Dev}&\textbf{MAD}\\
\midrule
Orbit 1  &$10^{-4}$& 2.71&0.0011&0.0010\\
Orbit 2  &$2\times10^{-4}$& 5.86&0.0108&0.0086\\
Orbit 3  &$1.9741\times10^{-3}$&6.15& 0.1559& 0.1344\\
Orbit 4  &$3\times10^{-3}$&6.14&0.0562&0.0452\\
\bottomrule
\end{tabular}
\end{table}

\begin{figure*}[htbp]
\center{
\includegraphics[scale=0.2]{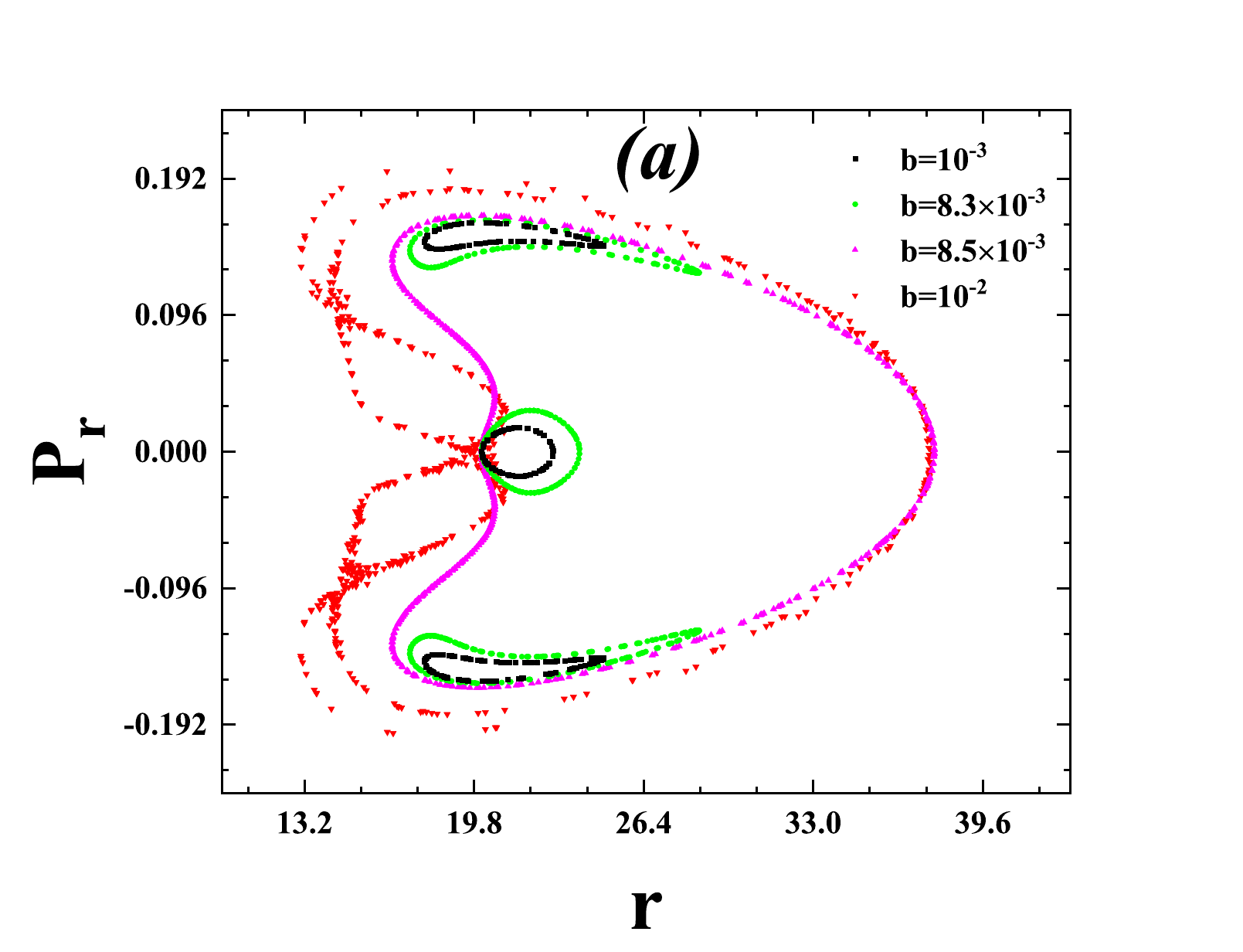}
\includegraphics[scale=0.2]{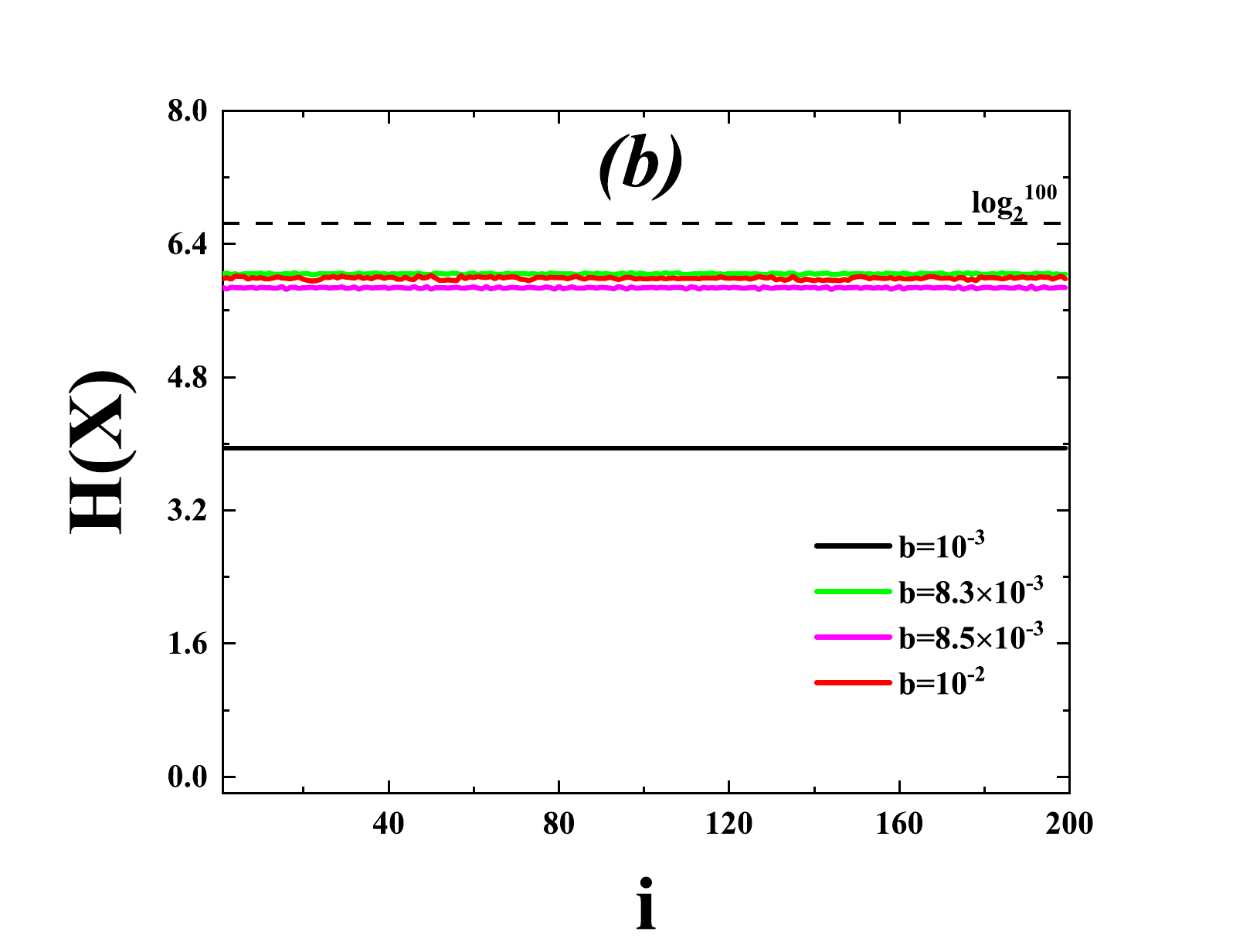}
\includegraphics[scale=0.2]{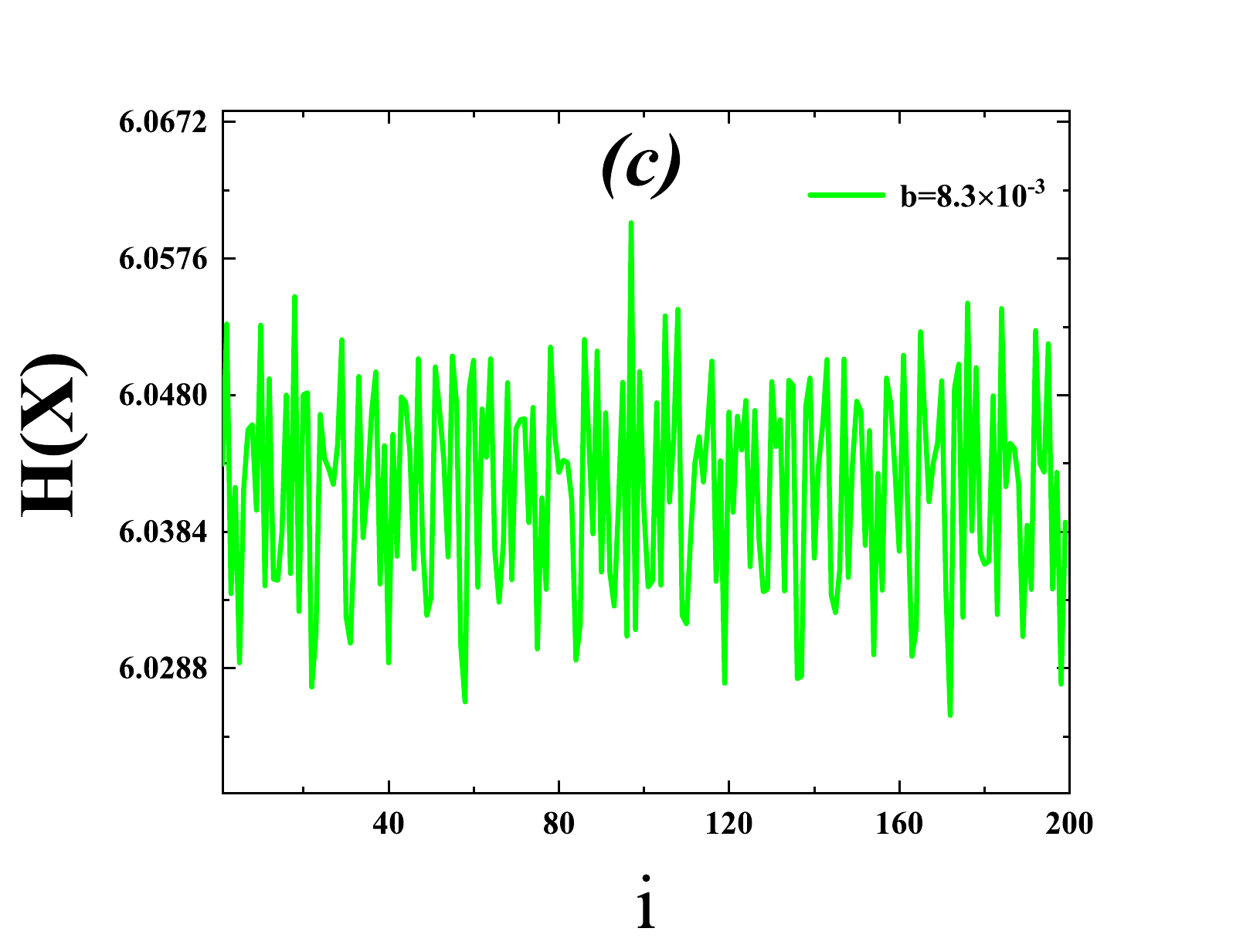}
\includegraphics[scale=0.2]{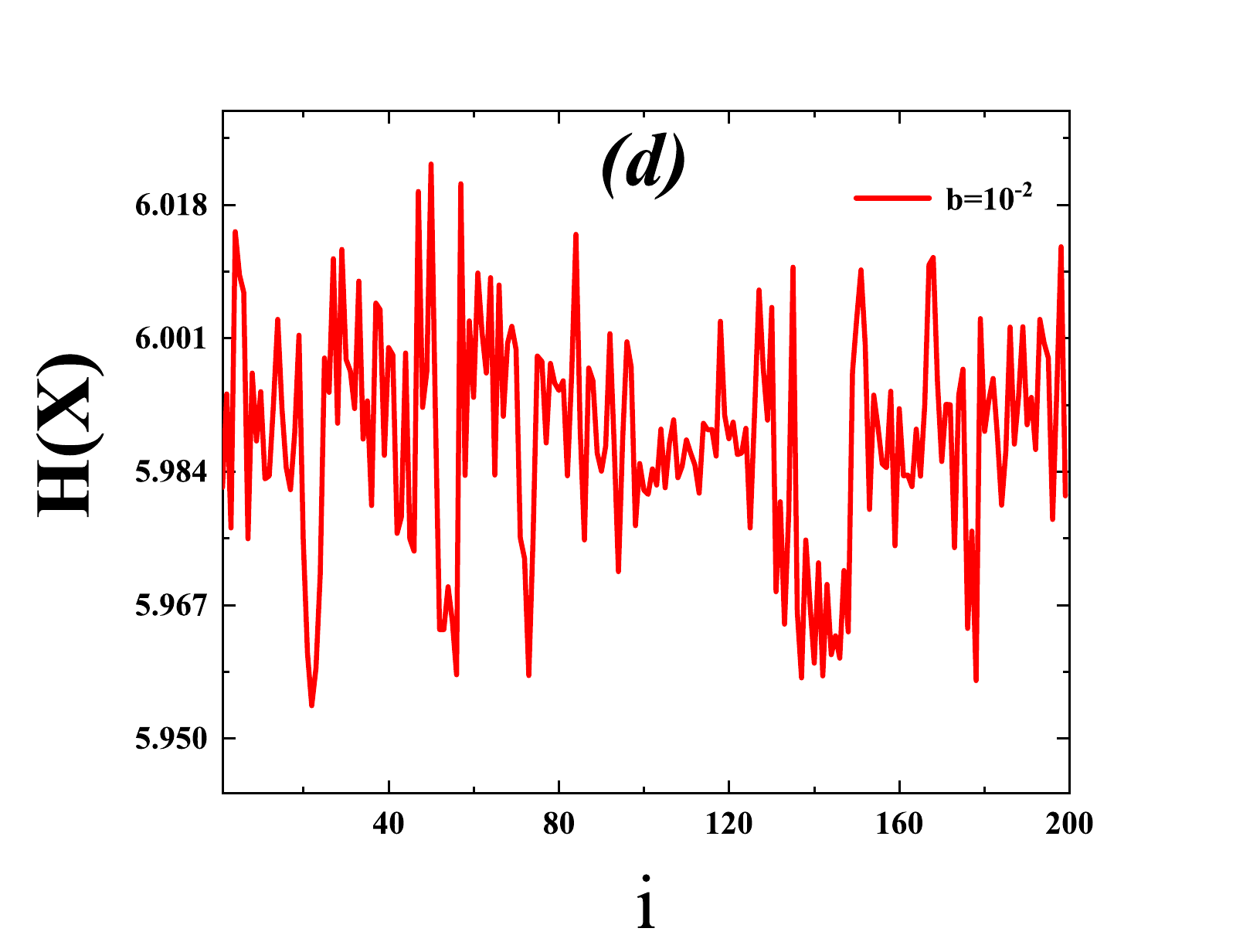}
\includegraphics[scale=0.3]{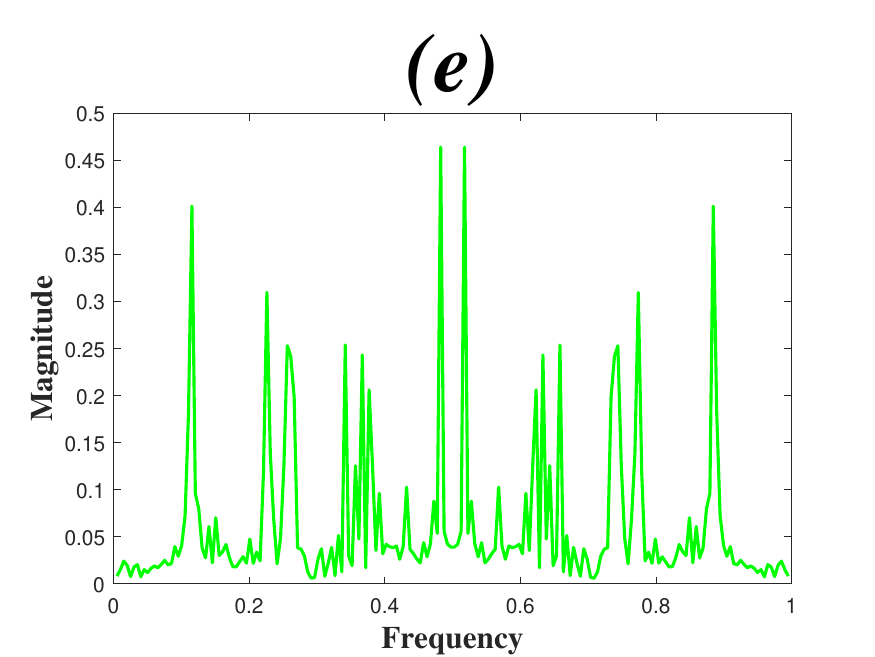}
\includegraphics[scale=0.3]{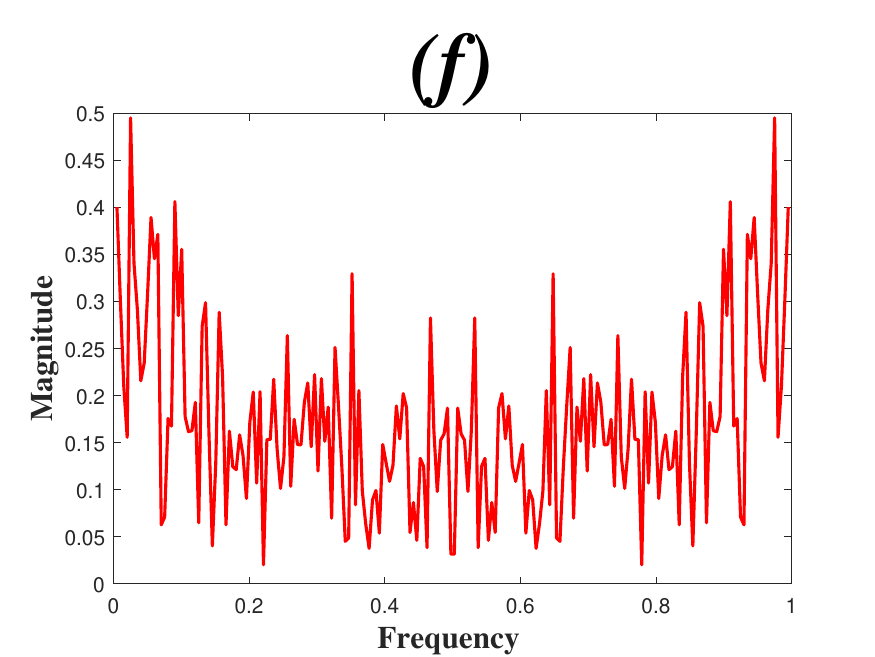}
\caption{The Poincar\'{e} map and Shannon entropy of charged particles moving around a black hole in an external magnetic field.
The initial position of the particle release is $c_{13}=0$, $c_{14}=0.9$, $r=20.103$, $\theta=\frac{\pi}{2}$, $E=0.976$, and $L=4.8$.
(a): The Poincar\'{e} section of charged particles as the magnetic field strengthens.
(b): The Shannon entropy of charged particles as the magnetic field strengthens.
(c)-(f): Shannon entropy of two types of orbits and their fast Fourier analysis in Fig. 6(b), with colors corresponding to those in the same figure.}
 \label{Fig5}}
\end{figure*}

\begin{table}[htbp]
\centering
\caption{Fluctuation Analysis of Orbital Entropy in Figure 6}
\begin{tabular}{lrrrrl}
\toprule
\textbf{Orbital ID}&$b$&\textbf{Mean}&\textbf{Std Dev}&\textbf{MAD}\\
\midrule
Orbit 1  &$10^{-3}$& 3.94&0.0003&0.0003\\
Orbit 2  &$8.3\times10^{-3}$&6.04&0.0072&0.0062\\
Orbit 3  &$8.5\times10^{-3}$&5.87& 0.0080&  0.0060\\
Orbit 4  &$10^{-2}$&5.98&0.0139& 0.0109\\
\bottomrule
\end{tabular}
\end{table}

\begin{figure*}[htbp]
\center{
\includegraphics[scale=0.2]{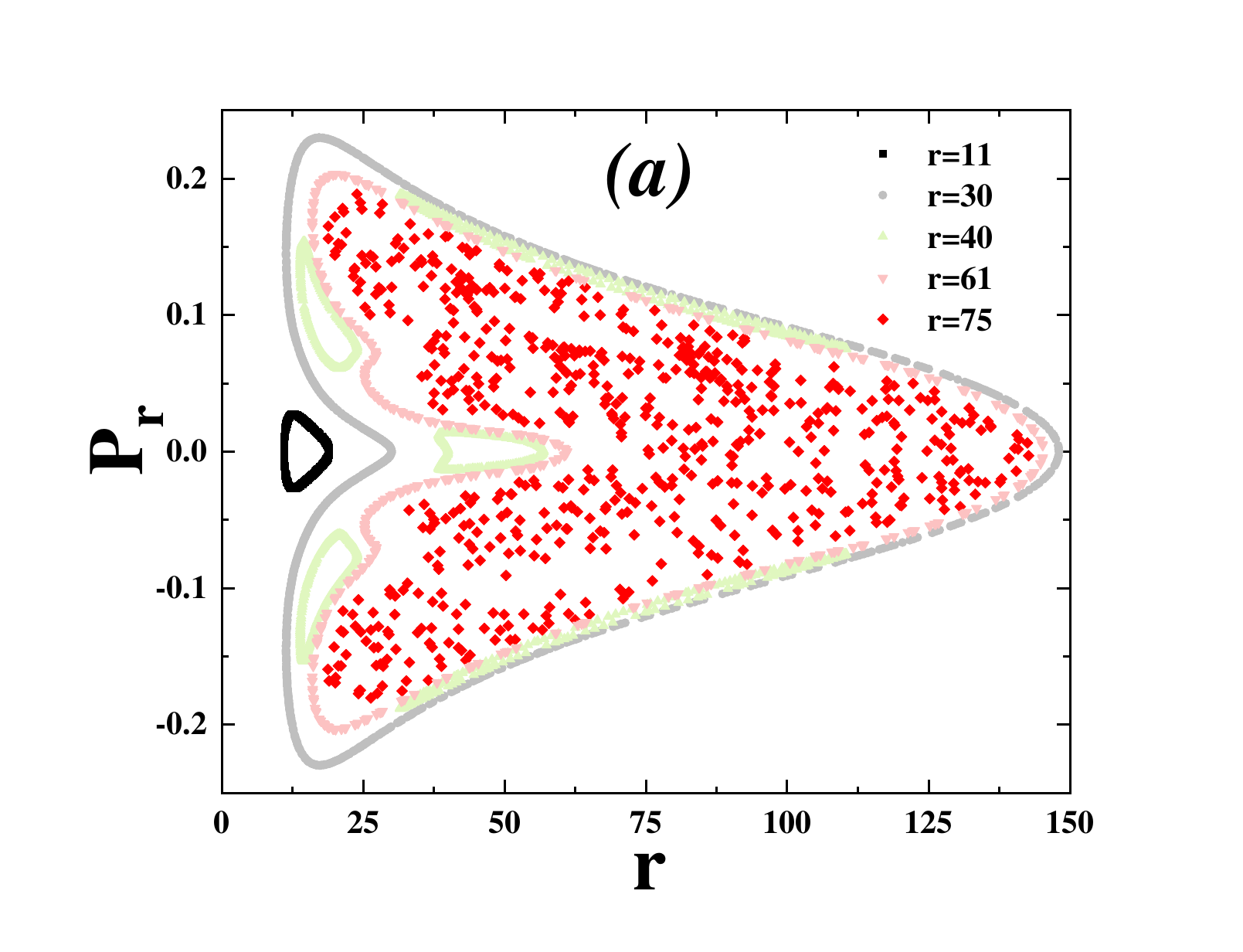}
\includegraphics[scale=0.2]{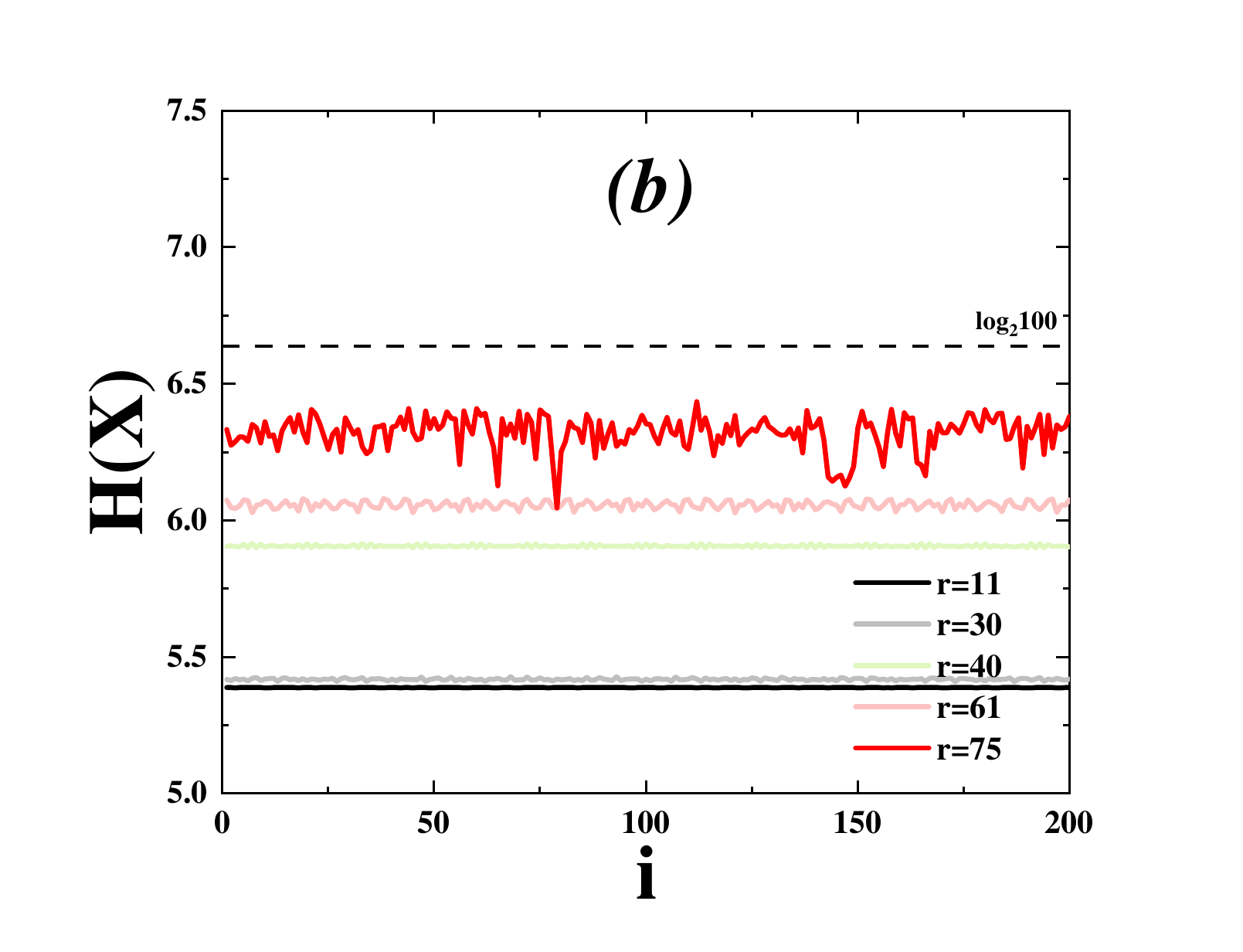}
\caption{The Poincar\'{e} map and Shannon entropy of charged particles moving around Kerr black hole in an external magnetic field.
The orbital parameters are: $E=0.995$, $L=4.6$, $a=0.5$, $b=0.001$, and $\theta=\frac{\pi}{2}$. The initial release positions of Orbits 1 to 5 are 11, 30, 40, 61, and 75, respectively.}
 \label{Fig5}}
\end{figure*}

\begin{table}[htbp]
\centering
\caption{Fluctuation Analysis of Orbital Entropy in Figure 7}
\begin{tabular}{lrrrrl}
\toprule
\textbf{Orbital ID}&$r$&\textbf{Mean}&\textbf{Std Dev}&\textbf{MAD}\\
\midrule
Orbit 1  &11& 5.38 &0.0003&0.0003\\
Orbit 2  &30&5.41&0.0041&0.0032\\
Orbit 3  &40&5.91& 0.0046&  0.0036\\
Orbit 4  &61&6.06&0.0145& 0.0122\\
Orbit 5  &75&6.32&0.0632& 0.0471\\
\bottomrule
\end{tabular}
\end{table}

\end{document}